\documentclass[preprint,12pt]{elsarticle}
\usepackage[top=1truein,bottom=1truein,left=1truein,right=1truein]{geometry}

\graphicspath{{./pic}}

\usepackage{amsmath}
\usepackage{amssymb}
\usepackage{amsfonts}

\usepackage{hyperref}
\usepackage{url}
\usepackage{orcidlink}

\usepackage{here}
\usepackage{tikz}
\usepackage{bm}
\usepackage{setspace}
\usepackage{enumerate}
\usepackage{arydshln}
\usepackage{caption}
\usepackage{color}
\usepackage{appendix}
\usepackage{lineno}
\usepackage{ulem}

\hypersetup{%
  colorlinks=true,
  linkcolor=magenta,
  citecolor=blue,
  urlcolor=cyan
}

\journal {...}

\begin {document}
\begin{frontmatter}

\title {Fine-tuning physics-informed neural networks for cavity flows using coordinate transformation}

% Affiliations
\affiliation[inst1]{
  organization={Department of Mechanical Systems Engineering, Graduate School of Systems Design, Tokyo Metropolitan University},
  addressline={1-1 Minami-Osawa}, 
  city={Hachioji},
  prefecture={Tokyo},
  postcode={192-0397}, 
  country={Japan}
}
\affiliation[inst2]{
  organization={Department of Mechanical Engineering, School of Engineering, Institute of Science Tokyo},
  addressline={2-12-1 Ookayama}, 
  city={Meguro-ku},
  prefecture={Tokyo},
  postcode={152-8550}, 
  country={Japan}
}

\author[inst1,inst2]{
  Ryuta Takao\,
}
\author[inst2]{
  Satoshi Ii\,\orcidlink{0000-0002-5428-5385}\corref{cor1}
}

\ead{ii.s.148c@m.isct.ac.jp}
\cortext[cor1]{Corresponding author.}

% Abstract about 200-250 words
\begin {abstract}
  Physics-informed neural networks (PINNs) have attracted attention as an alternative approach to solve partial differential equations using a deep neural network (DNN). Their simplicity and capability allow them to solve inverse problems for many applications. Despite the versatility of PINNs, it remains challenging to reduce their training cost. Using a DNN pre-trained with an arbitrary dataset with transfer learning or fine-tuning is a potential solution. However, a pre-trained model using a different geometry and flow condition than the target may not produce suitable results. This paper proposes a fine-tuning approach for PINNs with coordinate transformation, modelling lid-driven cavity flows with various shapes. We formulate the inverse problem, where the reference data inside the domain and wall boundary conditions are given. A pre-trained PINN model with an arbitrary Reynolds number and shape is used to initialize a target DNN. To reconcile the reference shape with different targets, governing equations as a loss of the PINNs are given with coordinate transformation using a deformation gradient tensor. Numerical examples for various cavity flows with square, rectangular, shear deformed and inflated geometries demonstrate that the proposed fine-tuning approach improves the training convergence compared with a randomly-initialized model. A pre-trained model with a similar geometry to the target further increases training efficiency. These findings are useful for real-world applications such as modelling intra-aneurysmal blood flows in clinical use.
\end{abstract}

% Research highlights
\begin{highlights} 
  \item We develop a fine-tuning approach for PINNs modelling lid-driven cavity flows.
  \item Pre-training with same shape, different Reynolds number improves convergence.
  \item Coordinate transformation extends fine-tuning approach to various cavity geometries.
\end{highlights}


\end{frontmatter}

% \linenumbers

%%% Section  %%%
\section{Introduction}
Physics-informed neural networks (PINNs) \cite{Raissi2019} have attracted attention as an alternative approach to solve partial differential equations (PDEs) using a deep neural network (DNN). In standard PINNs for solving PDEs, coordinates and time are the input, and physical quantities are the output of the DNN. Thus, partial derivatives of the physical quantities with respect to the input which are used to evaluate PDE loss are easily calculated by an automatic differentiation \cite{Baydin2017} through the DNN. An advantage of this approach is that it requires only a small amount of training data because physics laws (given by PDEs) are implicit in the DNN.

The simplicity and capability of PINNs is conducive to solving inverse problems \cite{Raissi2019}. If reference data such as experiment or observation data are given in the analysis domain, the PINNs can derive an inverse problem using a loss function according to data mismatch. This capability of PINNs is favorable in many applications. For example, in the medical field, PINNs are applied to evaluate patient-specific hemodynamics of intra-aneurysmal flows using observed low-resolution velocity data \cite{Raissi2020,Arzani2021,Moser2023,Zhang2023,Garay2024}.

\begin{figure}[H]
  \begin{center}
    \includegraphics[width=\linewidth]{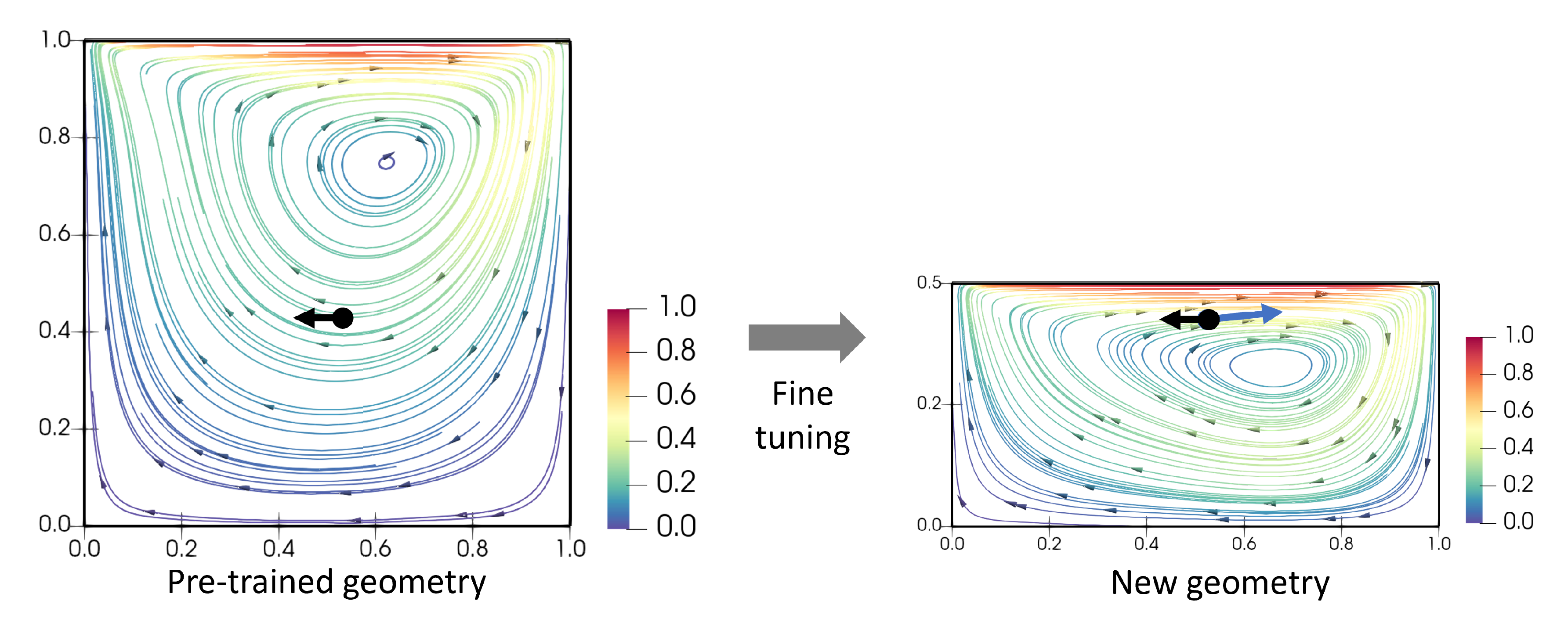}
    \caption{A difficulty of a fine-tuning approach that the pre-trained DNN is applied to a new DNN for a cavity flow with a different aspect ratio. The black arrows denote flow vectors at the same coordinate point in the reference configuration (square domain) and target configuration (rectangular domain), whereas the blue arrow indicates the correct flow vector in the target configuration.}
    \label{fig_s1_mot}
  \end{center}
\end{figure}

Despite the versatility of the PINNs, it remains challenging to reduce their training cost. PINNs entail a relatively long training time (e.g. \cite{Zhu2024}), hindering their practical usage compared with conventional numerical methods to solve PDEs. In this regard, using a pre-trained DNN model for an arbitrary dataset with transfer learning or fine-tuning is promising. However, extending to a different geometry poses a challenge. As shown in an example (Fig.~\ref{fig_s1_mot}), where a pre-trained model for a square cavity flow problem is used to initialize the DNN for a rectangular shape with a different aspect ratio, the fine-tuning approach is not straightforwardly applied. This is apparent because the velocity directions at a particular point are different between the square and rectangular shapes. A progressive boundary complexity approach (called boundary progressive PINN) \cite{Chen2025} or warm-start PINN \cite{Daneker2024}, which has been proposed to improve a training accuracy and efficiency by gradually changing the domain geometry from a reference shape to a target (complex) shape in the model training, may also be useful for the fine-tuning approach. However, a consistent strategy for setting hyper-parameters appropriately for the highly deformed shapes of real-world datasets (e.g., how gradually to change the shape, how many epochs to train each time, etc.) is not trivial.

In this work, we formulate a fine-tuning approach of PINNs for lid-driven cavity flows governed by the incompressible Navier-Stokes (NS) equations. The coordinate transformation is introduced with a deformation gradient tensor between the reference and target configurations, and the governing equations (PDEs of physics laws) given in the reference configuration are used for the PDE losses. The idea that the coordinate transformation (or geometric mapping/transformation) is embedded in the DNN has been seen in previous research \cite{Gao2021,Burbulla2023}; however, this concept has not yet been used to accomplish fine-tuning. We demonstrate the validity of the proposed fine-tuning approach with three numerical examples using synthetic data as a set of pseudo-observed velocities for 2D cavity flows with square, rectangular, shear deformed and inflated shapes.

%%% Section %%%
\section{Methods}
\subsection{Lid-driven cavity flow}
A lid-driven cavity flow is given by the stationary incompressible Navier-Stokes (NS) equations with a non-dimensional form:
\begin{align}
  & \nabla\cdot{\bf u} = 0,
  \label{eq:continuity} \\
  & {\bf u}\cdot\nabla{\bf u} = -\nabla{p} + \nabla\cdot{\bm \tau}, 
  \label{eq:momentum} \\
  & {\bm \tau} = \frac{1}{Re} (\nabla{\bf u} + \nabla{\bf u}^\top),
  \label{eq:viscous}
\end{align}
in spatial coordinate ${\bf x} \in \Omega \subset \mathbb{R}^d$, where $d$ is the dimension ($d=2$ or $3$), ${\bf u}({\bf x}) \in \mathbb{R}^d$ is the velocity vector, $p({\bf x}) \in \mathbb{R}$ is the pressure, $\nabla = \partial/\partial{\bf x} \in \mathbb{R}^d$ is the spatial derivative, and ${\bm \tau} \in \mathbb{R}^{d \times d}$ is the symmetric viscous stress tensor. $Re$ is the Reynolds number, which is defined by using the lid velocity $U_{lid}=1$, lid length $L_{lid}=1$ and kinematic viscosity $\nu$ as $Re=U_{lid}L_{lid}/\nu$. In lid-driven cavity flows, the above equations are solved with a no-slip boundary condition at the lid (or top) boundary $\Gamma_{lid}$ and the fixed-wall boundaries $\Gamma_{wall}$ of a cavity:
\begin{alignat}{2}
  & {\bf u} = U_{lid}{\bf e}_1, & \quad & {\rm for} \ {\bf x} \in \Gamma_{lid},
  \label{eq:bd_lid} \\
  & {\bf u} = 0, & \quad & {\rm for} \ {\bf x} \in \Gamma_{wall},
  \label{eq:bd_wall}
\end{alignat}
where ${\bf e}_1$ is the unit basis vector in the lid-driving direction $x_1$.

\subsection{PINNs in a current configuration}
We start with a standard formulation of PINNs for the stationary incompressible NS equations in a current configuration. A DNN with fully-connected layers is then introduced, consisting of an input layer ${\bf X}=\{{\bf x}\} \in \mathbb{R}^d$, output layer ${\bf Y}=\{{\bf u}, p, {\bm \tau}' \} \in \mathbb{R}^{(d+1)(d+2)/2}$, and hidden (or intermediate) layers ${\bf Z} = \{{\bf z}^1, {\bf z}^2, \cdots, {\bf z}^N\} \ \in \mathbb{R}^{N \times V}$, where $V$ is the number of neurons in the hidden layers (${\bf z}^n \in \mathbb{R}^V$ for $n \in [1,N]$), and ${\bm \tau}'=\{\tau_{ij} \ | \ i \leq j, \ i,j \in [1,d]\} \in \mathbb{R}^{d(d+1)/2}$ is the vector consisting of independent components of the symmetric tensor ${\bm \tau}$. Thus,
\begin{align}
  {\bf Y} = f({\bm \theta}; {\bf X})
  \ \Leftrightarrow \
  \begin{cases}
  {\bf z}^{0} = {\bf X}, \\
  {\bf z}^{n} = 
  \sigma\left({\bf W}^{n}{\bf z}^{n-1} + {\bf b}^{n}\right), 
    \quad (1 \leq n \leq N), \\  
  {\bf Y} = {\bf W}^{N+1}{\bf z}^{N} + {\bf b}^{N+1}.
  \end{cases}
  \label{eq:DNN}
\end{align}
Here, $f: {\bf X} \longmapsto {\bf Y}$ denotes the mapping through the DNN with the network parameters ${\bm \theta}=\{({\bf W}^1, {\bf b}^1), ({\bf W}^2,{\bf b}^2), \cdots, ({\bf W}^{N+1},{\bf b}^{N+1})\}$, where ${\bf W}^n \in \mathbb{R}^{\dim({\bf z}^{n-1}) \times \dim({\bf z}^n)}$ is the weight, ${\bf b}^n \in \mathbb{R}^{\dim({\bf z}^n)}$ is the bias, and $\sigma$ is the activation function, which is set to the $\tanh$ function in this study, i.e., $\sigma(x)=\tanh(x)$.

The parameters ${\bm \theta}$ are estimated by solving the following optimization problem:
\begin{align}
  {\bm \theta} = \mathop{\arg\min}\limits_{{\bm \theta}^*} \mathcal{L}({\bm \theta}^*),
  \label{eq:optimize}
\end{align}
where $\mathcal{L}({\bm \theta})$ is the total loss, which is a functional of the set of parameters ${\bm \theta}$, defined below. In the PINNs for inverse problems, the loss considers PDEs as physics laws (i.e., stationary incompressible NS equations in this study) and the reference data (generally observation/measurement data):
\begin{align}
  \mathcal{L} = \mathcal{L}_{PDE} + \lambda_{data}\mathcal{L}_{data}.
\end{align}
Here, $\mathcal{L}_{PDE}$ entails the PDE losses in terms of Eqs. \eqref{eq:continuity}, \eqref{eq:momentum}, and \eqref{eq:viscous}, and $\mathcal{L}_{data}$ is the data loss of the velocity mismatch with the reference velocity. The losses $\mathcal{L}_{PDE}$ and $\mathcal{L}_{data}$ are evaluated by a summation of $L_2$ norms of each component of a vector or second-order tensor. $\lambda_{data}$ is the loss weight for the PDE. The partial derivatives to evaluate $\mathcal{L}_{PDE}$ are calculated using automatic differentiation \cite{Baydin2017,Raissi2019}.

In this study, we enforce the boundary condition \eqref{eq:bd_wall} by the hard constraint manner \cite{Lu2021b}:
\begin{align}
  \mathcal{B}: \tilde{\bf u} \longmapsto {\bf u}, \quad {\rm for} \ \Gamma_{wall},
  \label{eq:hard_constraint}
\end{align}
where $\mathcal{B}$ is an arbitrary function mapping the velocity output from \eqref{eq:DNN} without satisfying boundary condition \eqref{eq:bd_wall}, $\tilde{\bf u}$, to that satisfying the boundary condition \eqref{eq:bd_wall}, ${\bf u}$. Note that, in this study, the boundary condition \eqref{eq:bd_lid} is not explicitly enforced on ${\bf u}$ (except for reference data defined at $\Gamma_{lid}$), and thus the system becomes an inverse problem that the velocity profile on $\Gamma_{lid}$ is automatically estimated from solving the optimization problem \eqref{eq:optimize} with the constraint \eqref{eq:hard_constraint}.

\begin{figure}[H]
  \begin{center}
    \includegraphics[width=\linewidth]{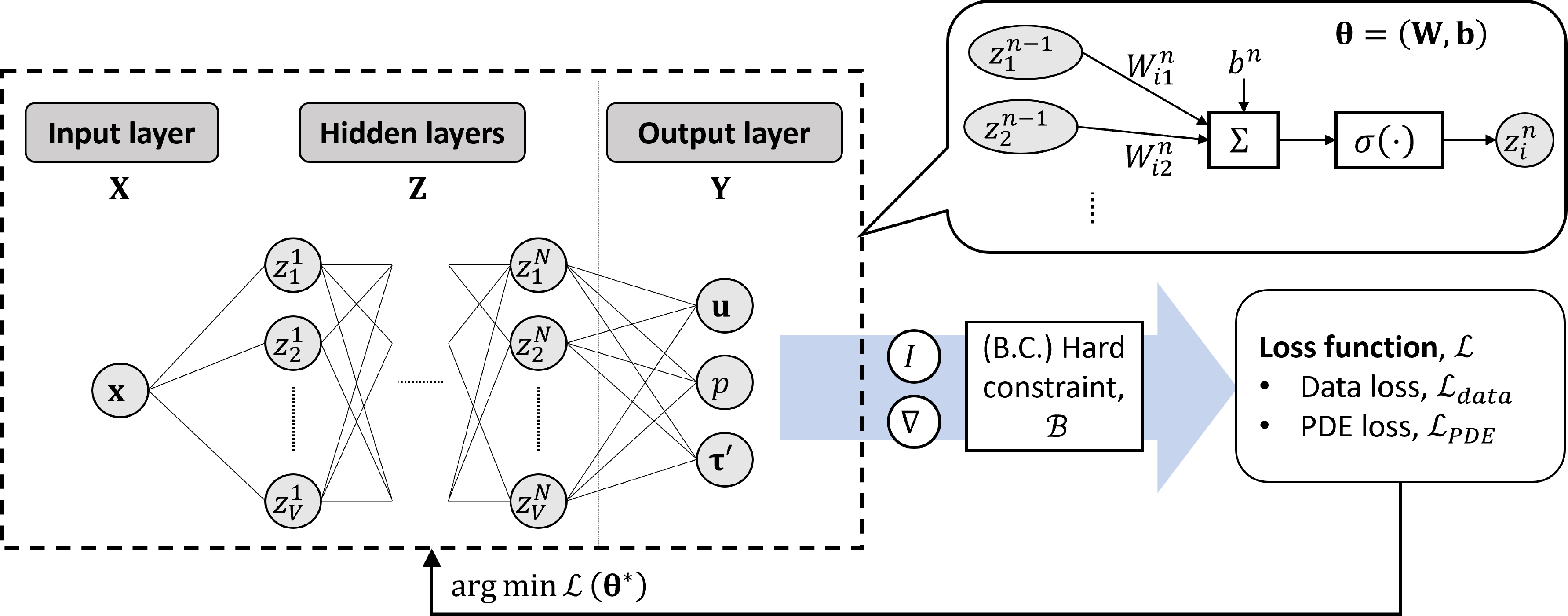}
    \caption{Overview of the PINN in the current configuration for the stationary incompressible NS equations.}
    \label{fig_s2_DNN}
  \end{center}
\end{figure}

It should be noted that the viscous term in Eq. \eqref{eq:momentum} can be simply given as $\nabla\cdot{\bm \tau} = Re^{-1}\nabla^2{\bf u}$ using Eqs.\eqref{eq:continuity} and \eqref{eq:viscous}, and in general this form is applied in standard PINNs for a single fluid flow without involving ${\bm \tau}$ at the output layer. However, when introducing the coordinate transformation described later, it is convenient to introduce the viscous stress tensor ${\bm \tau}$ to circumvent the difficulty of evaluating second-order derivatives of the coordinate Jacobian. 

\subsection{PINNs with coordinate transformation in a reference configuration}
\begin{figure}[H]
  \begin{center}
    \includegraphics[width=0.7\linewidth]{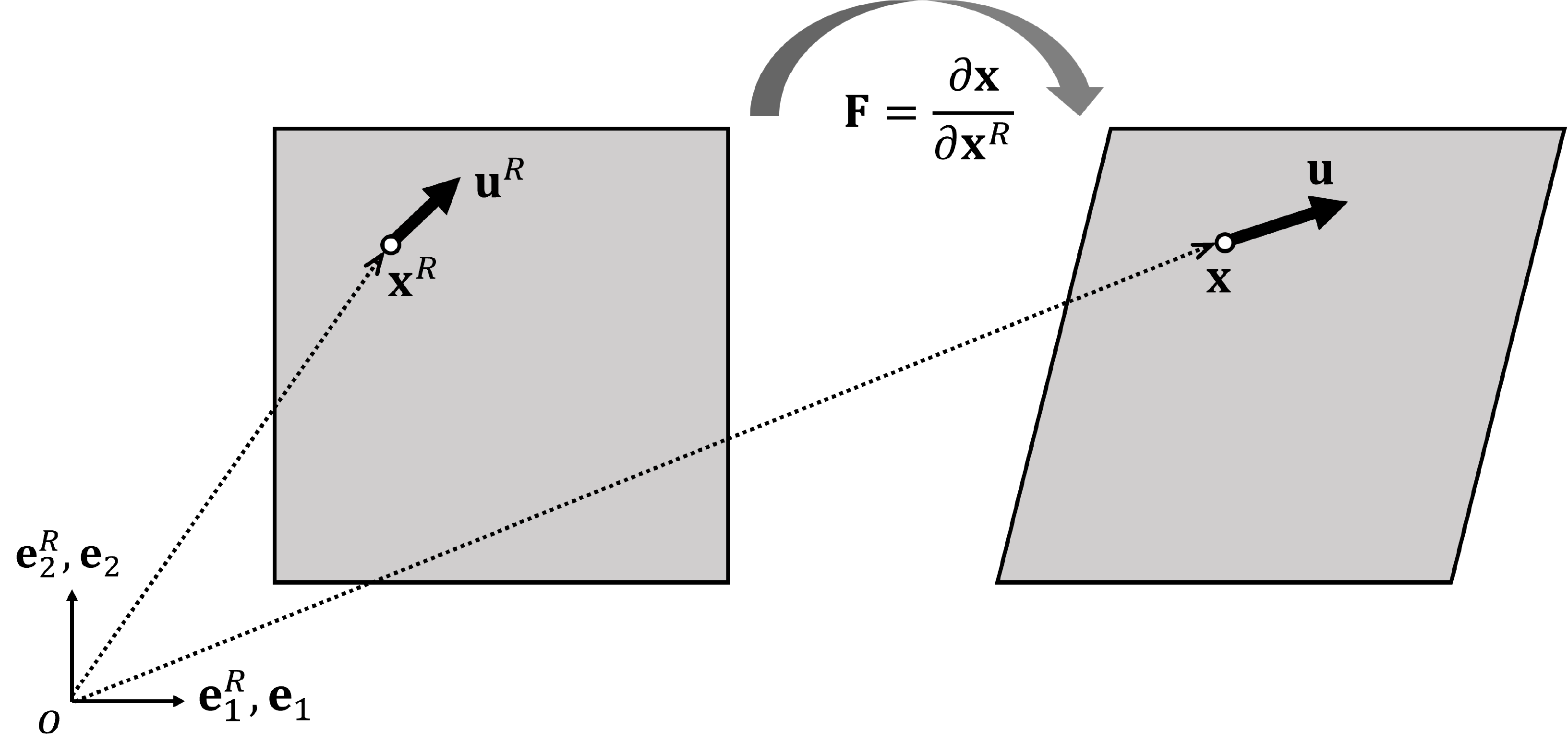}
    \caption{A schematic of the coordinate transformation between the reference configuration ${\bf x}^R=x^R_i{\bf e}^R_i$ and target configuration ${\bf x}=x_i{\bf e}_i$ ($i\in[1,d]$).}
    \label{fig_s2_deform}
  \end{center}
\end{figure}

Let ${\bf F}=\partial {\bf x}/\partial {\bf x}^R$ denote the deformation gradient tensor, where ${\bf x}^R \in \mathbb{R}^d$ and ${\bf x} \in \mathbb{R}^d$ denote coordinates in a reference (un-deformed) configuration and target (current) configuration, respectively (Fig.~\ref{fig_s2_deform}).

Using ${\bf F}$, the following relationships hold:
\begin{align}
  & {\bf u} = {\bf F}\cdot{\bf u}^R
  \ \Leftrightarrow \
  {\bf u}^R = {\bf F}^{-1}\cdot{\bf u}, \\
  & \nabla^R = {\bf F}^\top\cdot\nabla
  \ \Leftrightarrow \
  \nabla = ({\bf F}^\top)^{-1}\cdot\nabla^R,
\end{align}
where ${\bf u}^R$ and $\nabla^R$ are the velocity vector and spatial gradient in the reference configuration, respectively. Then, Eqs. \eqref{eq:continuity}, \eqref{eq:momentum}, and \eqref{eq:viscous} can be rewritten as:
\begin{align}
  & {\rm tr} ({\bf L}) = 0,
  \label{eq:t_continuity} \\
  & {\bf F}\cdot{\bf u}^R\cdot{\bf L} 
  = {\bf G}^\top\cdot(-\nabla^R{p} + \nabla^R\cdot{\bm \tau}), 
  \label{eq:t_momentum} \\
  & {\bm \tau} = \frac{1}{Re} ({\bf L} + {\bf L}^\top),
  \label{eq:t_viscous}
\end{align}
where ${\bf G}={\bf F}^{-1}$ and ${\bf L}=\nabla{\bf u}$ is the velocity gradient tensor described by
\begin{align}
  {\bf L} = {\bf G}^\top \cdot (\nabla^R{\bf u}^R\cdot{\bf F}^\top + \nabla^R{\bf F}\cdot{\bf u}^R).
  \label{eq:grad_vel}
\end{align}
The detailed description of each equation is shown in Appendix A.

\begin{figure}[H]
  \begin{center}
    \includegraphics[width=\linewidth]{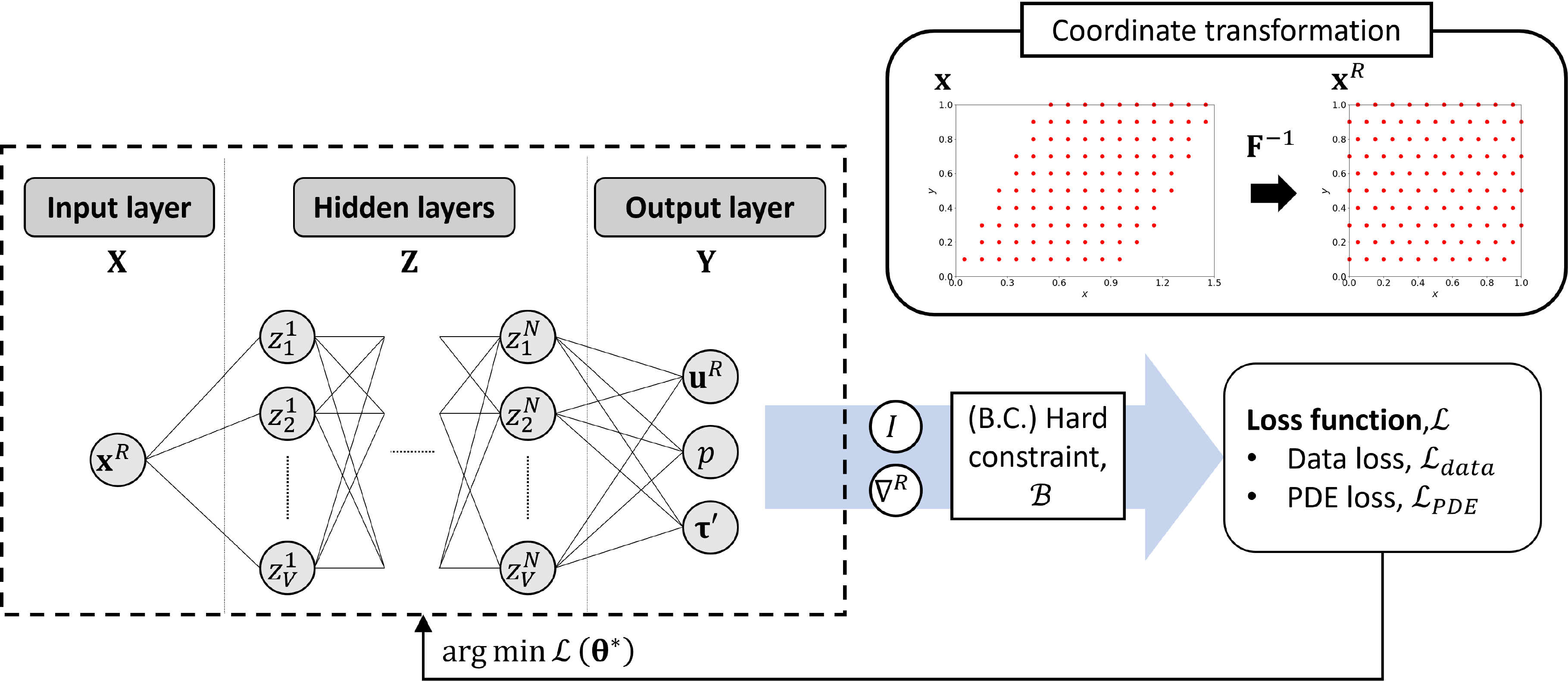}
    \caption{The proposed fine-tuning model for PINNs with coordinate transformation for cavity flows.}
    \label{fig_s2_DNN_ct}
  \end{center}
\end{figure}

In PINNs with coordinate transformations, the DNN is modified by replacing the input layer by ${\bf X}=\{{\bf x}^R\} \in \mathbb{R}^d$ and the output layer by ${\bf Y}=\{{\bf u}^R, p, {\bm \tau}' \} \in \mathbb{R}^{(d+1)(d+2)/2}$. Here, ${\bm \tau}'$ has the same definition as in the PINN without coordinate transformation, which consists of independent components of the symmetric tensor ${\bm \tau}$, thus, ${\bm \tau}'=\{\tau_{ij} \ | \ i \leq j, \ i,j \in [1,d]\} \in \mathbb{R}^{d(d+1)/2}$. The overview of the model is shown in Fig.~\ref{fig_s2_DNN_ct}. Analogous to the PINNs without coordinate transformation, the partial derivatives with respect ot ${\bf x}^R$, i.e., $\nabla^R$, can be calculated using automatic differentiation. Once the deformation gradient tensor ${\bf F}$ denoting the coordinate transformation is known, the modified DNN can evaluate Eqs. \eqref{eq:t_continuity}, \eqref{eq:t_momentum}, and \eqref{eq:t_viscous} using \eqref{eq:grad_vel}. Reader can see the expression of losses $\mathcal{L}_{data}$ and $\mathcal{L}_{PDE}$ in Appendix B.

In this study, we assume the reference geometry is set to a unit square, ${\bf x}^R=(x^R_1, x^R_2)=(x^R,y^R) \in [0,1]$, and the target geometry is given by a linear deformation with analytically set ${\bf F}$. We simply set the mapping function of the hard constraint of the Dirichlet boundary condition \eqref{eq:hard_constraint} in the unit square domain for the reference configuration as
\begin{align}
  {\bf u}^R = \mathcal{B}(\tilde{\bf u}^R) = x^R(1-x^R)y^R \tilde{\bf u}^R.
\end{align} 
\subsection{Synthetic training data}
We create the training data by numerically solving 2D lid-driven cavity flows with various geometries. The unsteady incompressible NS equations are solved with the boundary conditions \eqref{eq:bd_lid} and \eqref{eq:bd_wall} at $\Gamma_{wall}$. Here, we apply the Laplacian form of the viscous term, i.e., $Re^{-1}\nabla^2{\bf u}$. Steady-state solutions are obtained by OpenFOAM v9 (OpenCFD, UK) and used as reference data (and ground truth). The mesh resolution is set to 0.01.

\begin{figure}[H]
  \begin{center}
    \includegraphics[width=\linewidth]{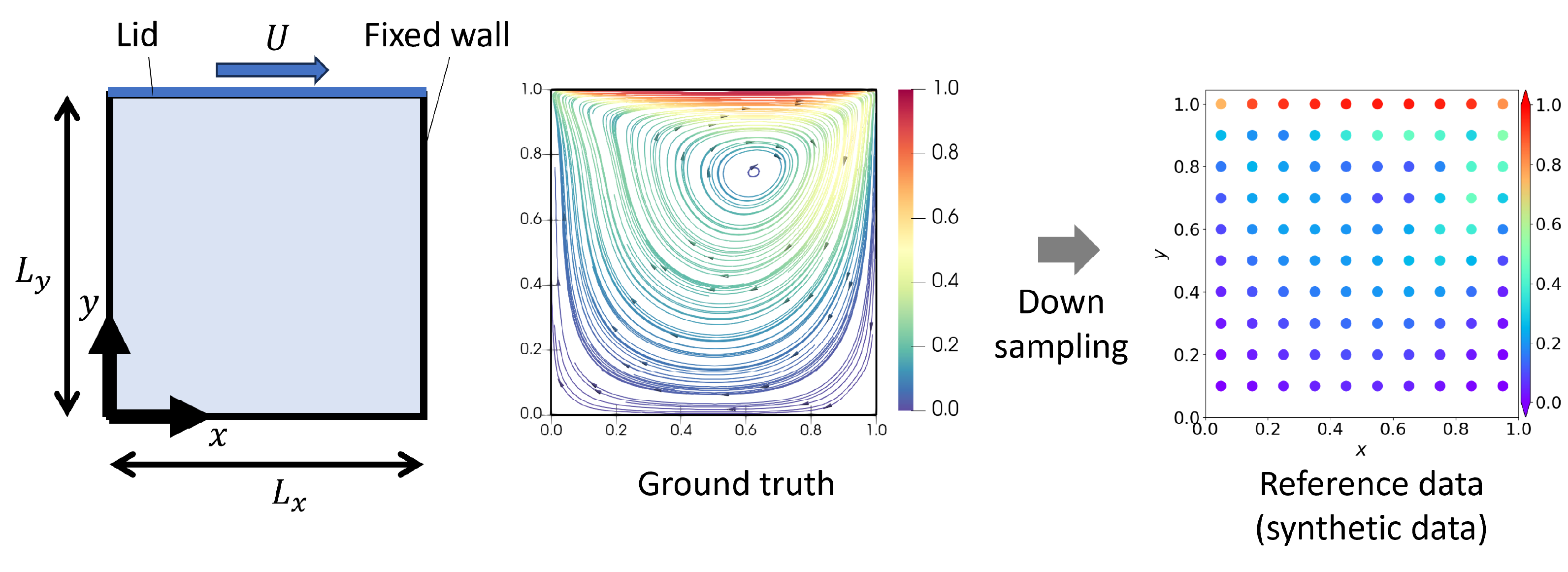}
    \caption{Construction of reference data (and ground truth) obtained by the computational fluid dynamics for two-dimensional cavity flow problems.}
    \label{fig_s2_openfoam}
  \end{center}
\end{figure}

\subsection{Training of the PINNs}
We use the DeepXDE \cite{Lu2021} Python library to implement the PINNs. The Adam optimizer \cite{Diederik2017} is used for model training, with a learning rate $\beta = 10^{-4}$ and the total number of epochs $5 \times 10^5$. The number of hidden layers and neurons in each hidden layer are set to $V=32$ and $N=4$, respectively. The loss weight is set to $\lambda_{data}=1$.

In this work, the fine-tuning model initializes the DNN using a model with different $Re$ or cavity shapes (termed pre-trained model), whereas the Glorot (Xavier) uniform initializer \cite{Glorot2010} is used for the pre-trained and randomly-initialized models. 

%%% Section %%%
\section{Results and discussion}
%%% Subsection %%%
\subsection{Square cavity flows}
\begin{figure}[H]
  \begin{center}
    \includegraphics[width=0.7\linewidth]{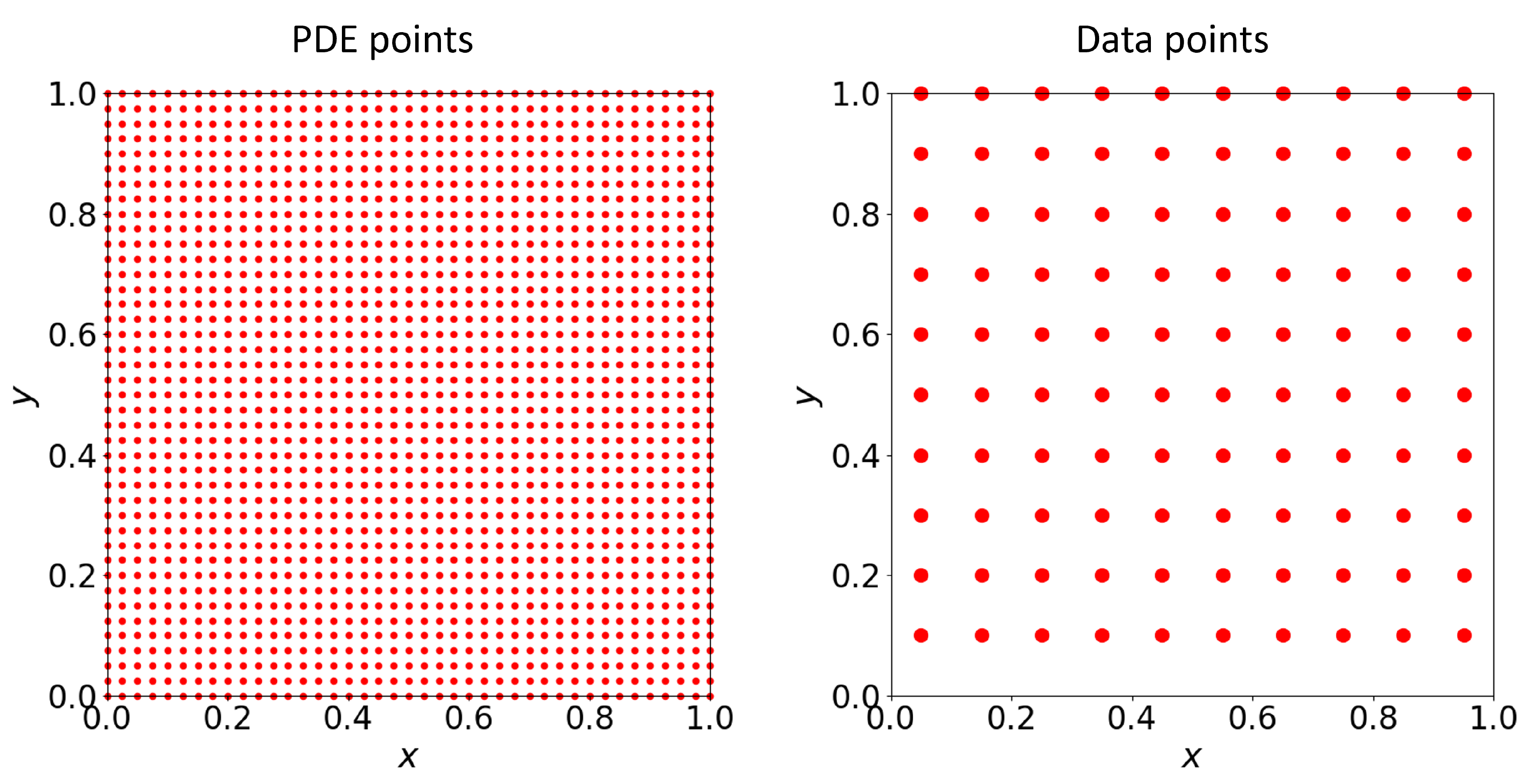}
    \caption{PDE points and data points, where the number of points are set to $N_{PDE}=1681$ and $N_{data}=100$, respectively.}
    \label{fig_s3_pts}
  \end{center}
\end{figure}

We use square cavity flows with a unit edge length. The sampling points for the PDE loss and data loss, termed PDE points and data points, respectively, are shown in Fig.~\ref{fig_s3_pts}. Also, to consider the case where measurement velocity is observed, the spatial resolution of the data points is set relatively low with respect to the domain size. The numbers of PDE points and data points are set to $N_{PDE}=1681$ (spatial resolution is 0.025) and $N_{data}=100$ (spatial resolution $h$ is 0.1), respectively.

\begin{figure}[H]
  \begin{center}
    \includegraphics[width=0.85\linewidth]{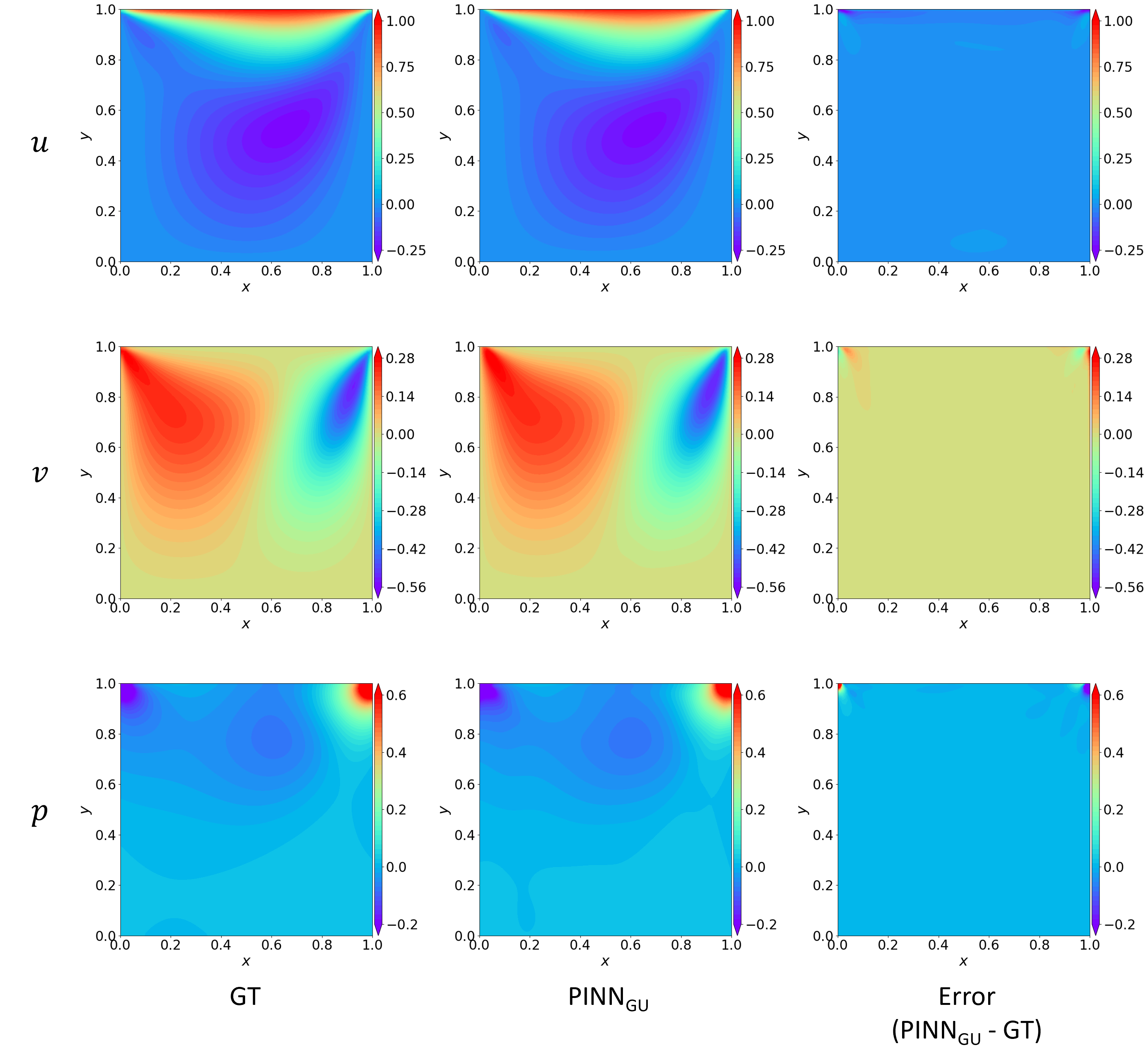}
    \caption{Comparisons between the ground truth (GT) and PINN with Glorot uniform initialization (PINN$_{GU}$) for the velocity $u$, $v$ and pressure $p$. The error for each is also shown.}
    \label{fig_s3_squ_color}
  \end{center}
\end{figure}

\begin{figure}[H]
  \begin{center}
    \includegraphics[width=\linewidth]{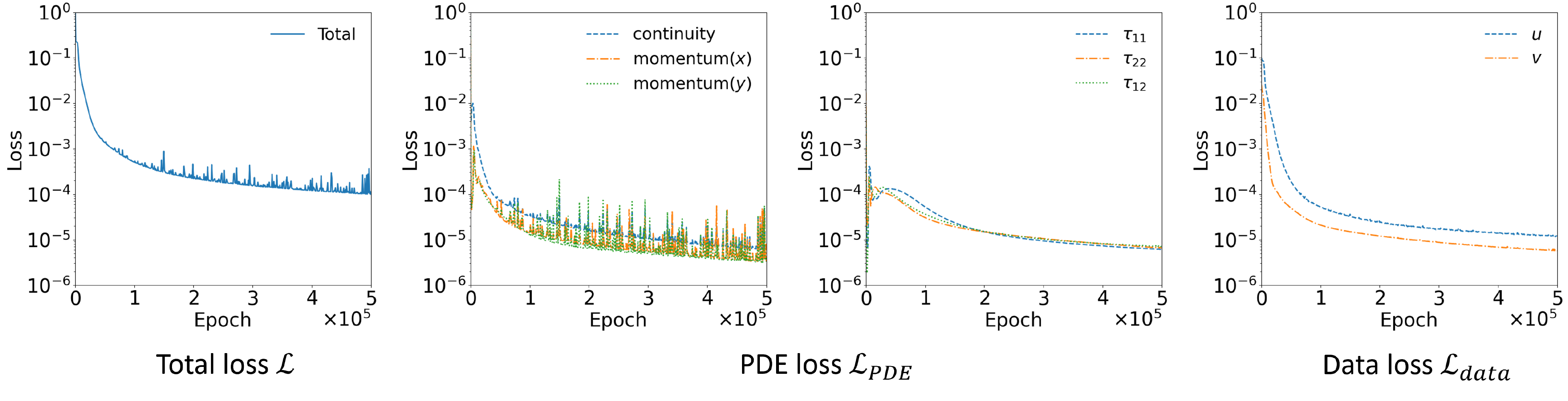}
    \caption{Respective loss histories in training, where ${\bf u}=(u,v)$ and ${\bm \tau}'=(\tau_{11},\tau_{22},\tau_{12})$.}
    \label{fig_s3_squ_loss}
  \end{center}
\end{figure}

First, the flow with $Re=100$ is solved. Fig.~\ref{fig_s3_squ_color} shows the comparisons between the ground truth and PINNs using the Glorot uniform initializer for the DNN, i.e., no pre-trained model is used, and Fig.~\ref{fig_s3_squ_loss} shows the histories of the training losses. These PINNs can reproduce the overall distributions of the ground truth for the velocity and pressure. Note that the average value of the pressure is set to $0$ for all the results. Some discrepancies occur around the top edge, especially in corners, where the boundary condition is not enforced because the lid velocity is inversely estimated in this example. Because numerical errors inevitably exist in both the PINNs and ground truth including discretization, spatial refinement, and boundary condition errors, the losses of the PINNs do not converge to $0$. While this approach does not achieve a flawless reproduction of the ground truth solution, the errors are confined near the top region, where the boundary velocity is inversely estimated. Thus, these PINNs adequately evaluate the overall flow profiles.


\begin{figure}[H]
  \begin{center}
    \includegraphics[width=\linewidth]{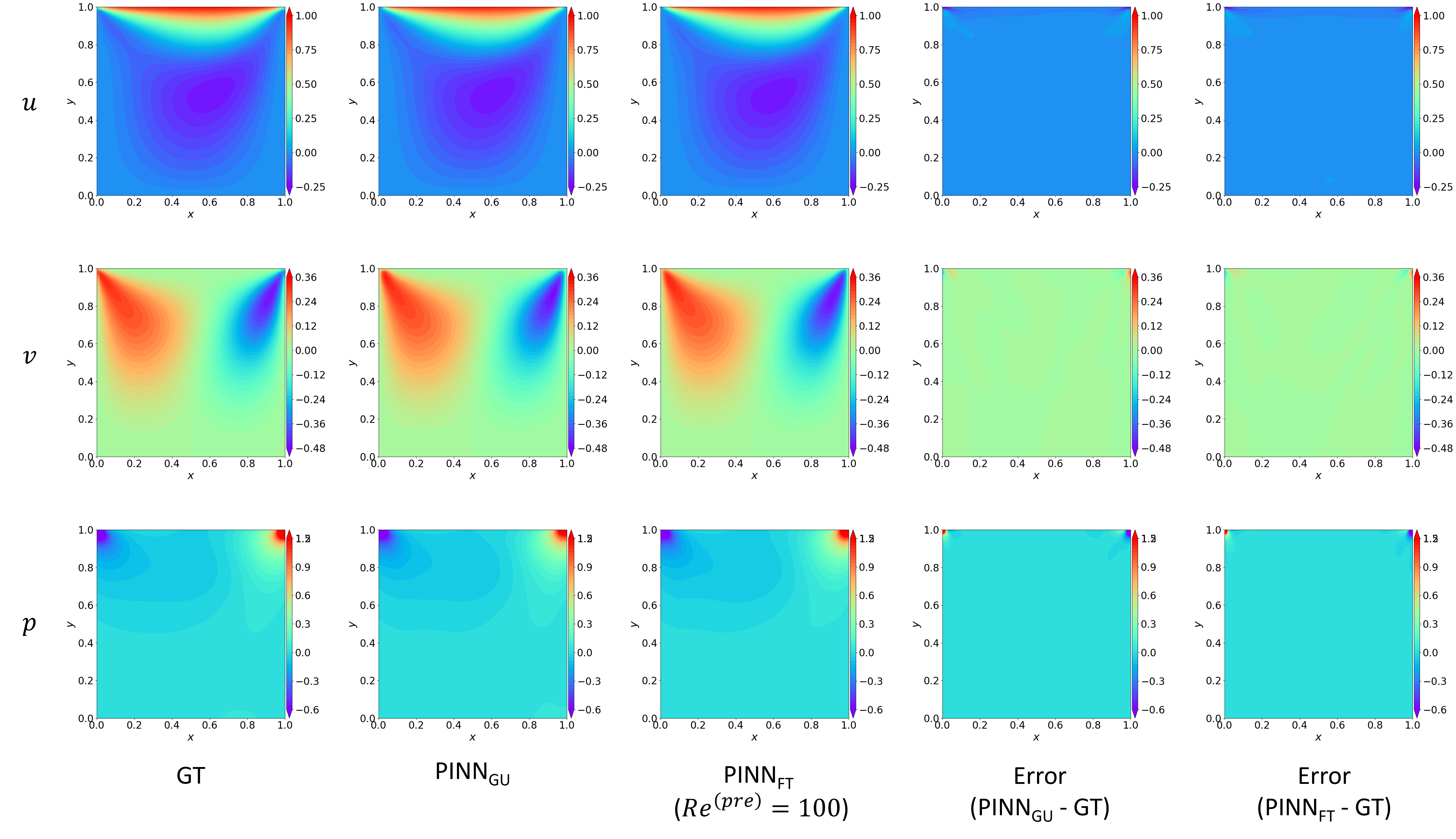}
    \caption{Comparisons for the ground truth (GT) and PINNs with Glorot uniform (PINN$_{GU}$) and fine-tuning (PINN$_{FT}$ with $Re^{(pre)}=100$) for the velocity $u$, $v$, and pressure $p$ at $Re=50$. The error for each is also shown.}
    \label{fig_squ_Re50}
  \end{center}
\end{figure}

\begin{figure}[H]
  \begin{center}
    \includegraphics[width=\linewidth]{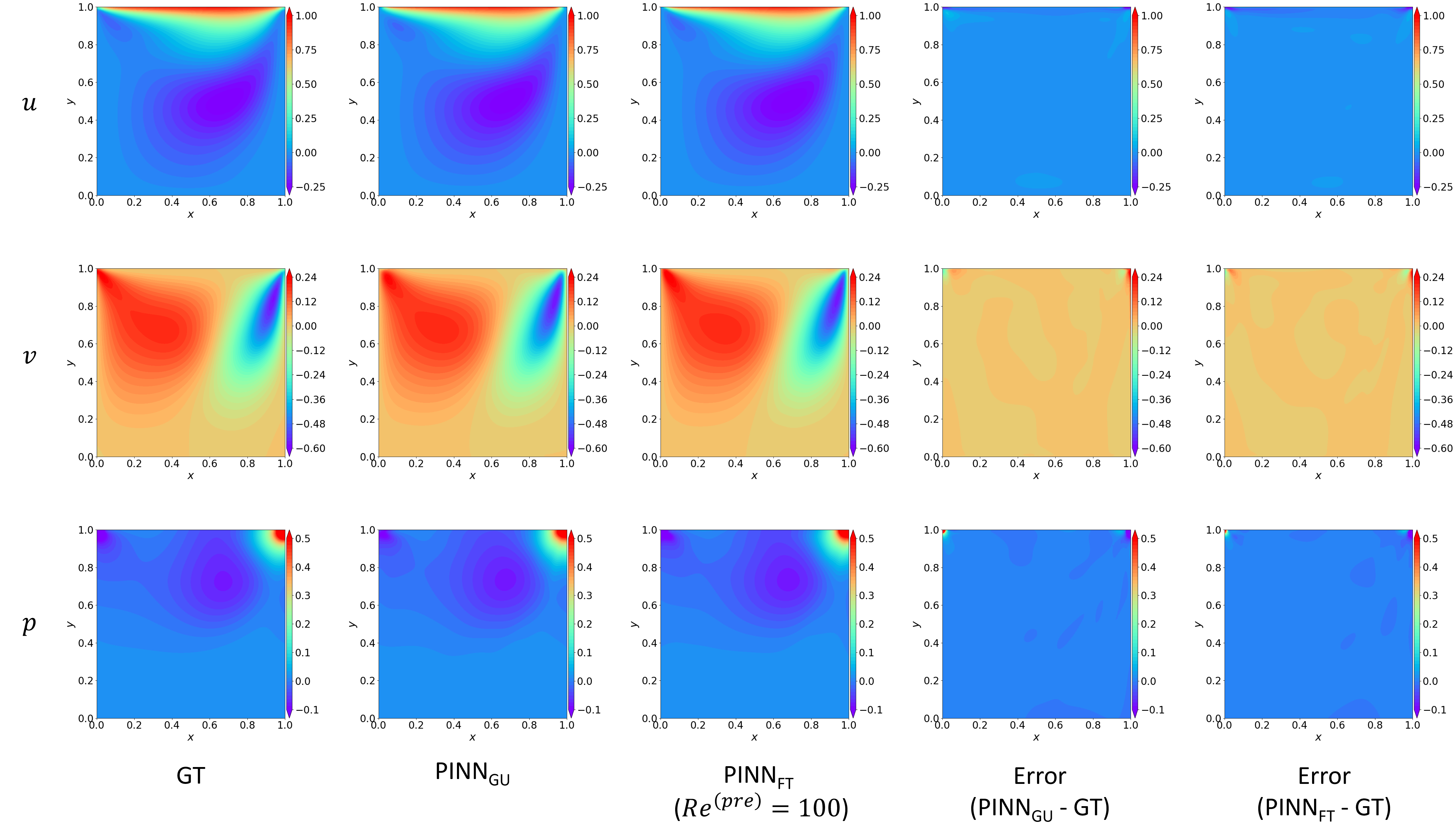}
    \caption{Comparisons for the ground truth (GT) and PINNs with Glorot uniform (PINN$_{GU}$) and fine-tuning (PINN$_{FT}$ with $Re^{(pre)}=100$) for the velocity $u$, $v$, and pressure $p$ at $Re=200$. The error for each is also shown.}
    \label{fig_squ_Re200}
  \end{center}
\end{figure}

To evaluate the performance of the fine-tuning approach, different $Re$ are applied. The PINNs with $Re=100$ (shown in the previous example) are used as the pre-trained model in the fine-tuning, and the training is performed for the flows at $Re=50$ and $200$. Fig.~\ref{fig_squ_Re50} and Fig.~\ref{fig_squ_Re200} show the flow fields of the ground truth and PINNs with Glorot uniform (PINN$_{GU}$) and fine-tuning (PINN$_{FT}$) at $Re=50$ and $200$, respectively. The PINNs capture the ground-truth flow fields with small error overall, except in the upper corner regions.


\begin{figure}[H]
  \begin{center}
    \includegraphics[width=0.7\linewidth]{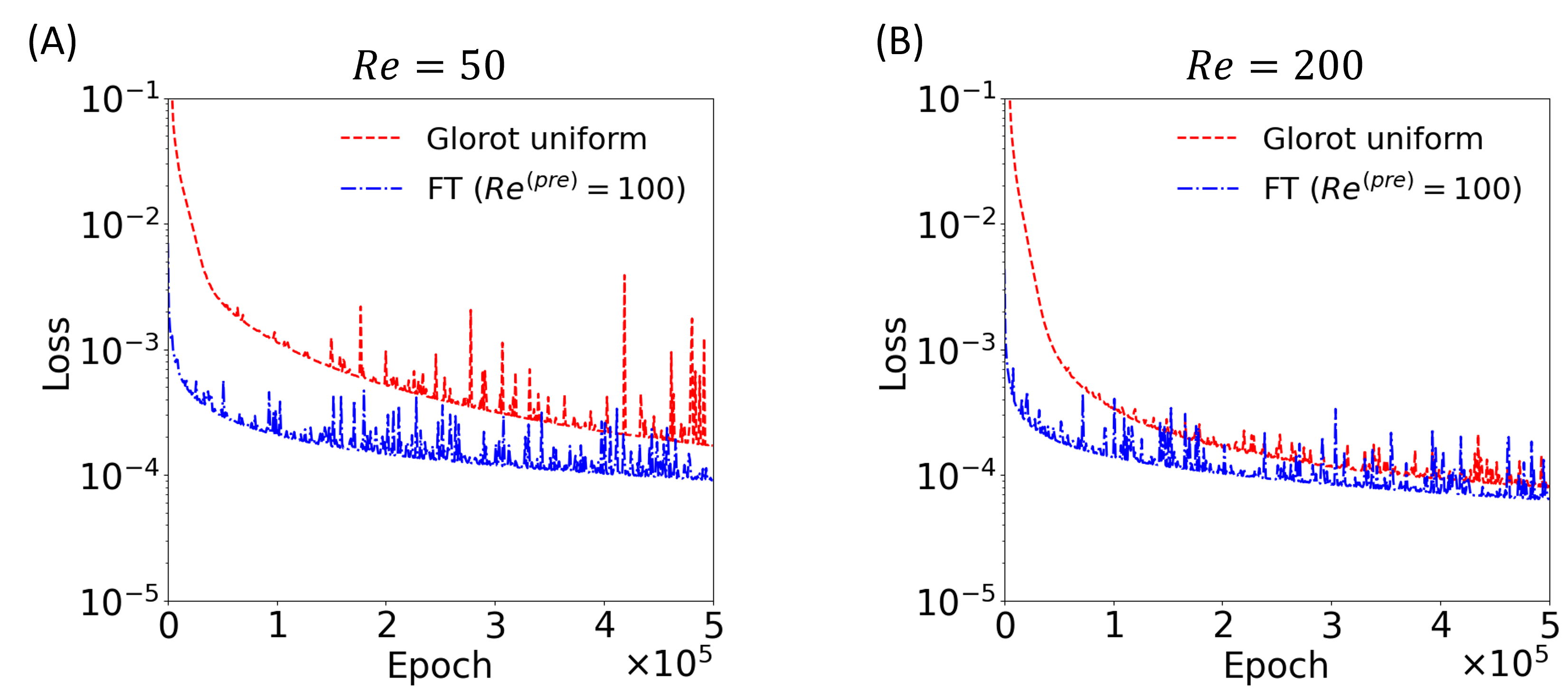}
    \caption{Convergence behaviors of the randomly-initialized model (i.e., Glorot uniform initialization) and fine-tuning model (FT) using the pre-trained model with $Re^{(pre)}=100$ for $Re=50$ (A) and $Re=200$ (B) in the square cavity flows.}
    \label{fig_s3_squ_ft}
  \end{center}
\end{figure}

Fig.~\ref{fig_s3_squ_ft} compares convergence behaviors of the losses between the randomly-initialized model (i.e., Glorot uniform initialization) and fine-tuning model. In both cases for $Re=50$ and $200$, the fine-tuning approach much improves the training convergence, such that the total loss is rapidly decreased compared with the randomly-initialized model. The cavity flows at low $Re$ demonstrate the common flow behavior where an overall circulation occurs from the driving lid, followed by small opposite circulations near bottom corners as $Re$ moderately increases. Therefore, the present fine-tuning approach, which uses a pre-trained DNN as the initialization of DNN, effectively reduces the number of learning epochs required to obtain the same level of loss. Evaluating higher $Re$ flows is more challenging with respect to formulating the PINNs, and further investigation is needed to confirm the behavior of the present fine-tuning approach for these cases.

%%% Subsection %%%
\subsection{Rectangular cavity flows with different aspect ratios}
We investigate the effects of the fine-tuning approach on cavity flows in several rectangular domains with different aspect ratios. The rectangular shape is defined by $1 \times \varepsilon$, where $\varepsilon$ is the aspect ratio of the cavity. Thus, the deformation gradient tensor is defined as:
\begin{align}
  {\bf F} = 
  \begin{bmatrix}
    1 & 0 \\
    0 & \varepsilon
  \end{bmatrix}.
\end{align}
The PINN with the square domain $\varepsilon=1$ is used as the pre-trained DNN (as shown in the section 3.1) for the fine-tuning model. In this example, the Reynolds number is fixed to $Re=100$.

\begin{figure}[H]
  \begin{center}
    \includegraphics[width=\linewidth]{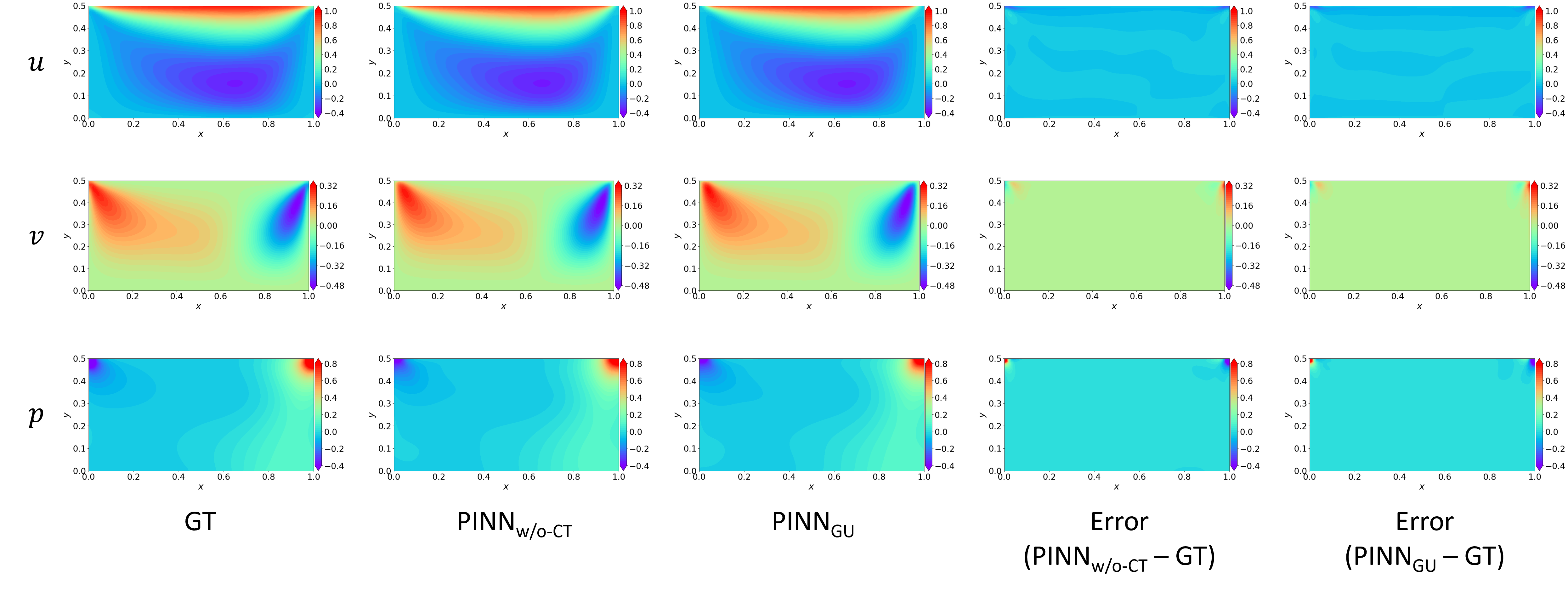}
    \caption{Comparisons of the flow fields for the ground truth (GT), PINN without coordinate transformation (PINN$_{w/o-CT}$) and PINN with Glorot uniform (PINN$_{GU}$) for the velocity $u$, $v$, and pressure $p$ at $\varepsilon=0.5$. The error for each is also shown.}
    \label{fig_rect_eps0.5}
  \end{center}
\end{figure}

First, we confirm the impact of the applied coordinate transformation on the estimation accuracy by comparing it with the PINN without coordinate transformation (PINN$_{w/o-CT}$). For this reason, we apply the PINN with the current configuration (Section 2.2) to the target rectangular cavity. In this case, the mapping of the hard constraint is given by ${\bf u}=\mathcal{B}(\tilde{\bf u})=x(1-x)y\tilde{\bf u}$. Fig.~\ref{fig_rect_eps0.5} shows the flow fields for the ground truth, PINN$_{w/o-CT}$ and PINN with Glorot uniform (PINN$_{GU}$) at $\varepsilon=0.5$.

\begin{figure}[H]
  \begin{center}
    \includegraphics[width=0.7\linewidth]{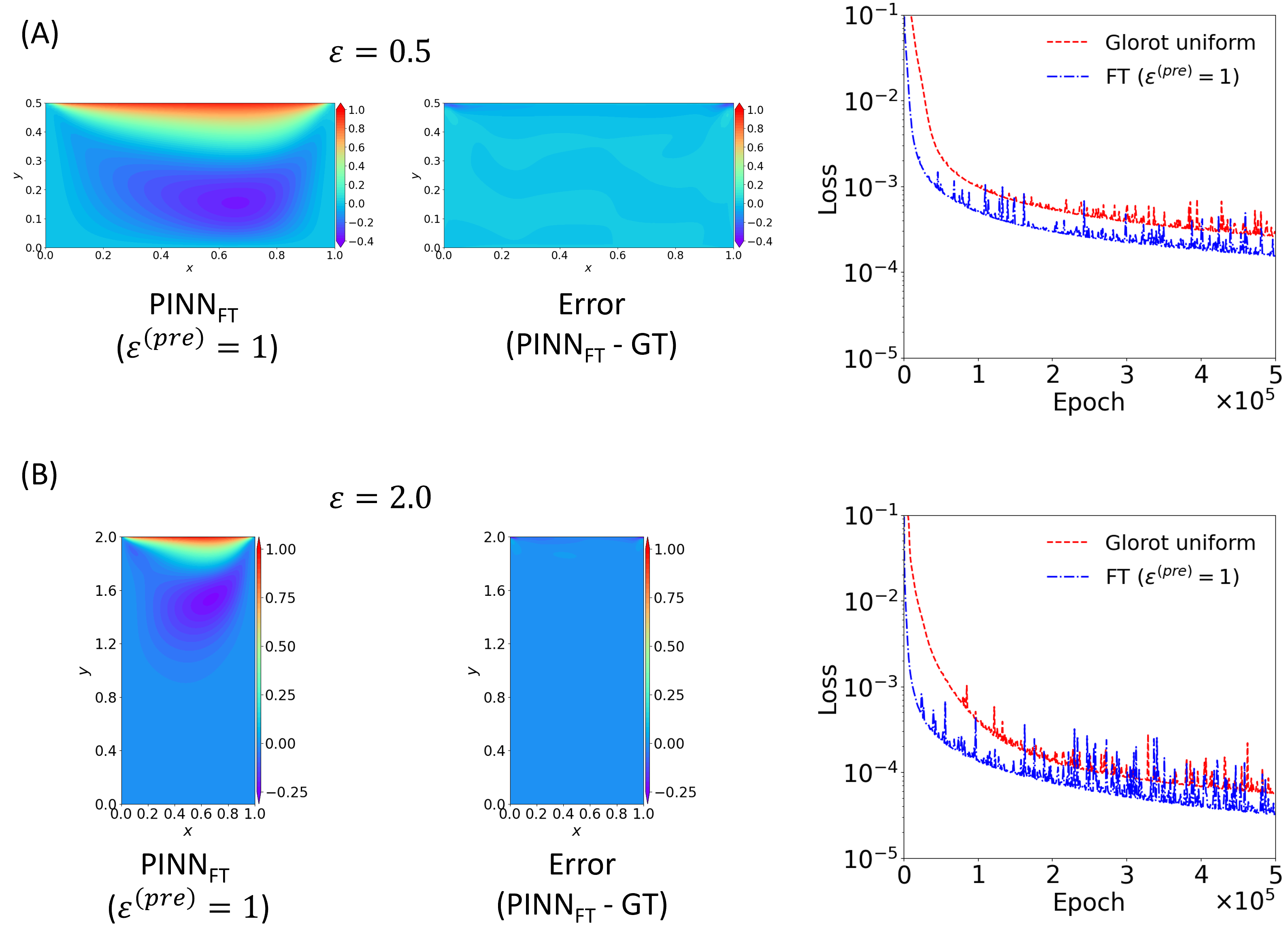}
    \caption{The fine-tuning results with the pre-trained model of $\varepsilon^{(pre)}=1$ (PINN$_{FT}$) with for the rectangular cavity shapes at $\varepsilon=0.5$ (A) and $\varepsilon=2$ (B) with respect to the velocity $u$ and the error to the randomly-initialized model (Glorot uniform: GT) as well as convergence behaviors of the total losses.}
    \label{fig_s3_rect_comp}
  \end{center}
\end{figure}

The fine-tuning results (PINN$_{FT}$) with $\varepsilon^{(pre)}=1$ for $\varepsilon=0.5$ and $2$ are shown in Fig.~\ref{fig_s3_rect_comp}, along with comparisons with the randomly-initialized model with respect to the convergence behaviors of the total losses. The velocity distributions obtained by the fine-tuning approach correspond closely to the reference, and the losses converge. Compared with the randomly-initialized model, the fine-tuning approach reduces the number of epochs to obtain the same loss level.

\begin{figure}[H]
  \begin{center}
    \includegraphics[width=\linewidth]{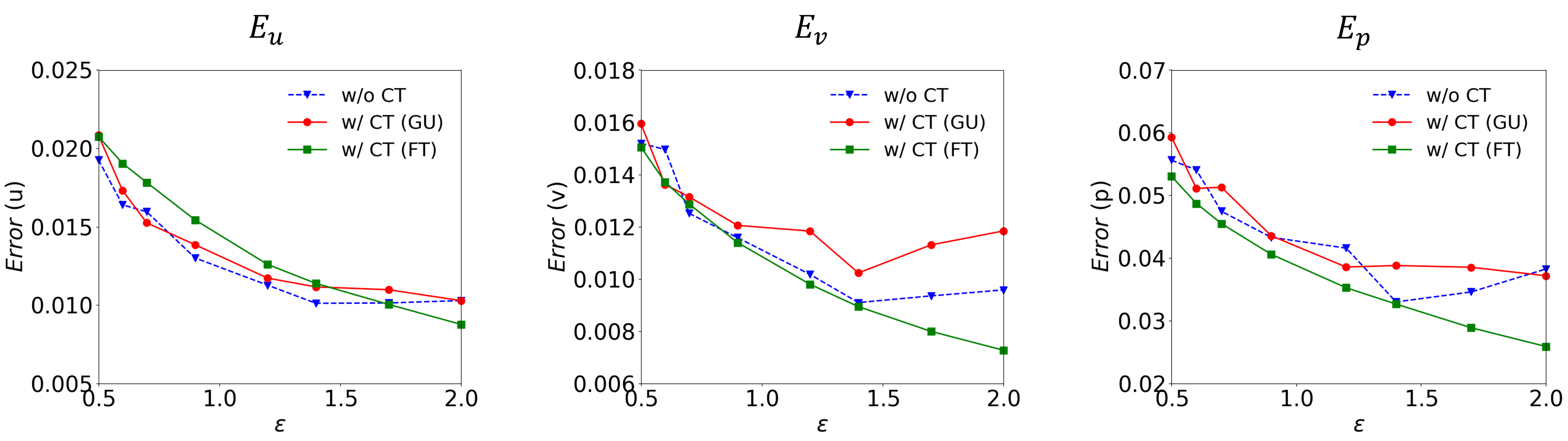}
    \caption{Comparisons of the $L_2$-norm errors of the velocity and pressure ($E_u, E_v, E_p$) for different rectangular shapes ($\varepsilon$) between the ground truth and PINNs without/with coordinate transformation (CT), where `w/o CT', `w/ CT (GU)' and `w/ CT (FT)' denote the PINN$_{w/o-CT}$ (without CT), PINN$_{GU}$ (with CT, Glorot uniform initializer) and PINN$_{FT}$ (with CT, fine-tuned with $\varepsilon^{(pre)}=1$), respectively.}
    \label{fig_rect_err}
  \end{center}
\end{figure}

Fig.~\ref{fig_rect_err} shows comparisons of the errors between the ground truth and each PINN (PINN$_{w/o-CT}$, PINN$_{GU}$, and PINN$_{FT}$ with $\varepsilon^{(pre)}=1$) to investigate the impact of coordinate transformation on estimation accuracy. The error is given by the $L_2$ norm of the discrete data at the PDE points. The error is relatively small compared to the absolute value shown in Fig.~\ref{fig_rect_eps0.5}. The overall behavior is that the error decreases with the shape change due to increasing $\varepsilon$. Importantly, there is no obvious difference in the error for each type of PINN, indicating that coordinate transformation does not degrade estimation accuracy.

To investigate how the learning process in the fine-tuning approach is changed according to the target cavity geometry, we performed further trials with various $\varepsilon$. The following index is introduced to evaluate convergence efficiency:
\begin{align}
  R_K = \frac{K_{FT}}{K_{UT}},
\end{align}
where $K_{FT}$ and $K_{UT}$ are the epochs of the fine-tuned and randomly-initialized models at the point when the total loss $\mathcal{L}$ falls below a threshold of $2 \mathcal{L}^{(UT)}_{min}$. Here, $\mathcal{L}^{(UT)}_{min} = \min \mathcal{L}^{(UT)}$ is the local minimum obtained in the randomly-initialized model. Namely, $R_K < 1$ signifies that the learning convergence is improved. 

\begin{figure}[H]
  \begin{center}
    \includegraphics[width=\linewidth]{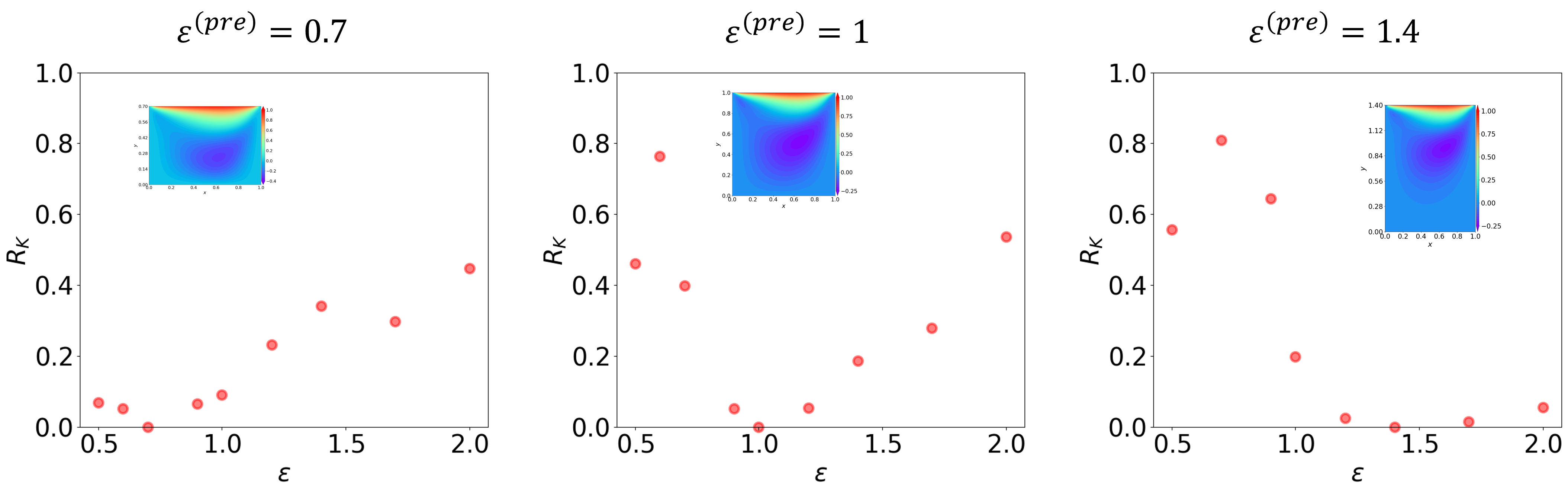}
    \caption{Relationship between $\varepsilon \in [0.5, 2]$ and $R_K$ when different shapes with $\varepsilon^{(pre)}=0.7$ (left), $1$ (middle), and $1.4$ (right) are used to pre-train the DNN model in the fine-tunings.}
    \label{fig_s3_rect_conv}
  \end{center}
\end{figure}

Fig.~\ref{fig_s3_rect_conv} shows the relationship between $\varepsilon \in [0.5, 2]$ and $R_K$ when the cavity shapes at $\varepsilon^{(pre)}=0.7$, 1 and 1.4 are applied for the pre-trained models for the fine-tuning. Here, in the case of own $\varepsilon$, the model has already converged, and thus $R_K=0$ is achieved. The results clearly show a convergence improvement for all cases and that the minimum value is shifted to the $\varepsilon$ value applied in the pre-trained model. These results suggest that it is preferable to use a similar shape in pre-training the DNN to improve the convergence behavior in the fine-tuning approach.

%%% Subsection %%%
\subsection{Cavity flows with shear-deformed shapes}
The next example investigates cavity flows with shear-deformed shapes:
\begin{align}
  {\bf F} = 
  \begin{bmatrix}
    1 & F_{12} \\
    0 & 1
  \end{bmatrix}.
\end{align}
In this test, we set $F_{12}$ to a constant and $Re=100$ for the flows.

\begin{figure}[H]
  \begin{center}
    \includegraphics[width=\linewidth]{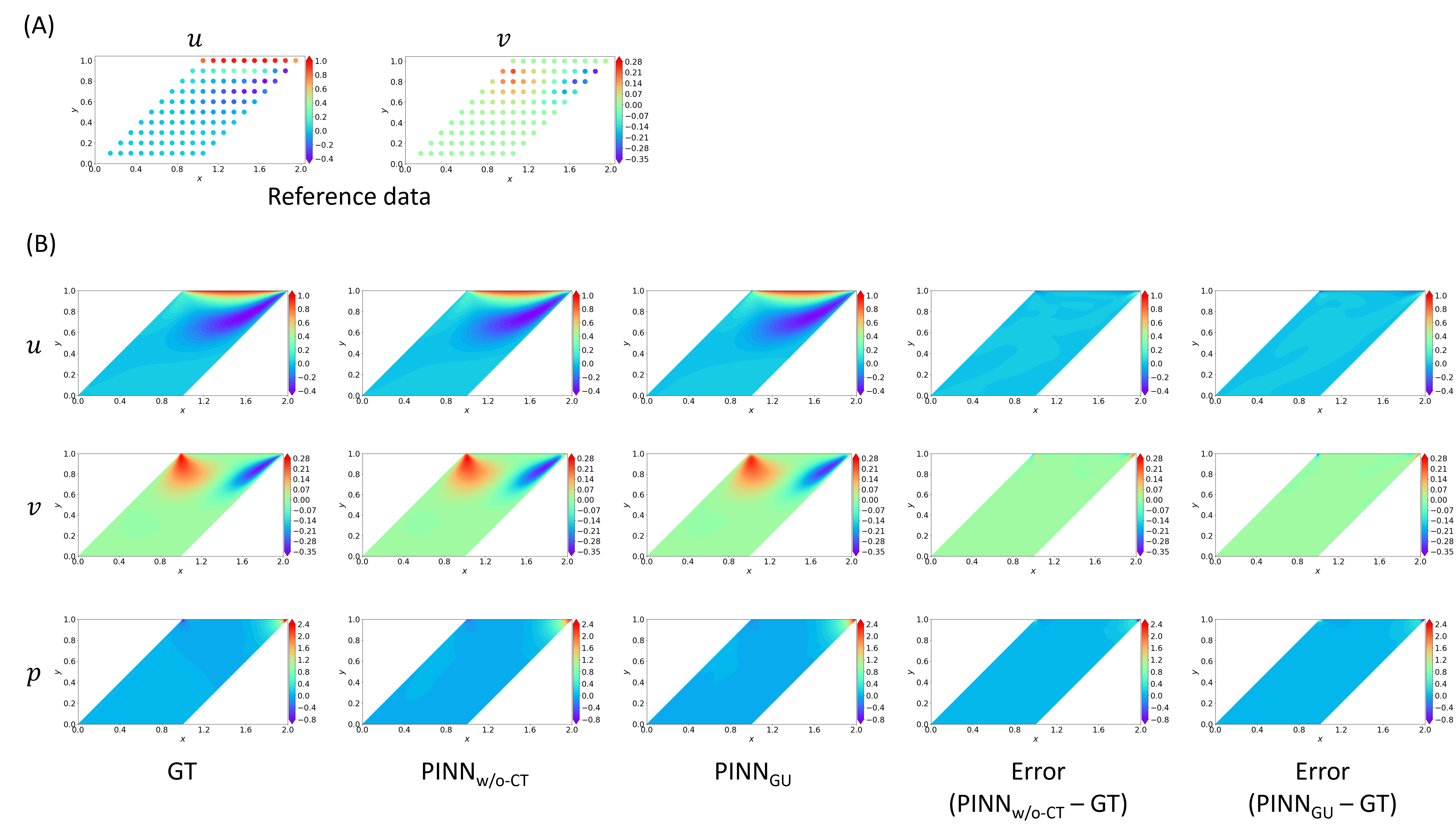}
    \caption{The reference data (A) and flow fields (B) among the ground truth (GT) and PINNs withtout and with coordinate transformation using the Glorot uniform initialize (PINN$_{w/o-CT}$ and PINN$_{GU}$) for $F_{12}=1$.}
    \label{fig_s3_def_comp}
  \end{center}
\end{figure}

\begin{figure}[H]
  \begin{center}
    \includegraphics[width=\linewidth]{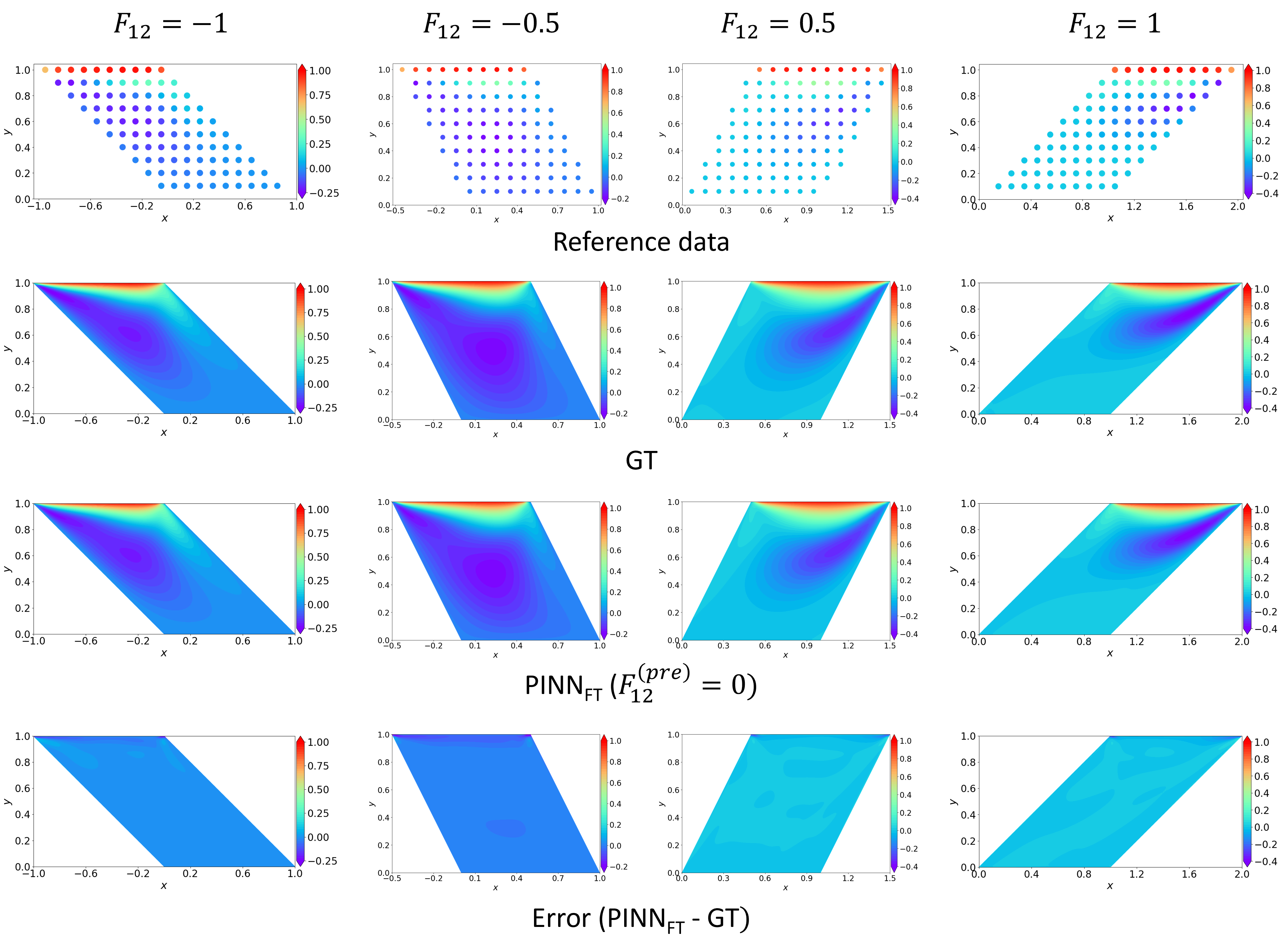}
    \caption{The results of velocity $u$ for $F_{12}=\{-1, -0.5, 0.5, 1\}$ are summarized, where the reference data, ground truth (GT) and fine-tuning model (PINN$_{FT}$) with $F_{12}^{(pre)}=0$ as well as their differences (errors) are shown.}
    \label{fig_s3_def_vF}
  \end{center}
\end{figure}

Fig.~\ref{fig_s3_def_comp} shows the ground truth and PINN flow fields without and with coordinate transformation using Glorot uniform initialization (PINN$_{w/o-CT}$ and PINN$_{GU}$) for $F_{12}=1$ to investigate the impact of the applied coordinate transformation on the estimation accuracy. Again, PINN$_{w/o-CT}$ is based on the target shear-deformed cavity in the current configuration (Section 2.2). Here, the hard-constraint mapping is given by ${\bf u}=\mathcal{B}(\tilde{\bf u})=(x+F_{12}y)(1-x-F_{12}y)y\tilde{\bf u}$. Regardless of the coordinate transformation, both models adequately capture the reference velocity and pressure distributions, except for the lid region. Fig.~\ref{fig_s3_def_vF} shows the results of fine-tuning for the velocity $u$ in the shear geometry for $F_{12}=\{-1, -0.5, 0.5, 1\}$ using a pre-trained model obtained under the condition $F_{12}^{(pre)}=0$ (square domain). The velocity profile of each ground truth is suitably reproduced using a few reference data for different cavity shapes.


\begin{figure}[H]
  \begin{center}
    \includegraphics[width=\linewidth]{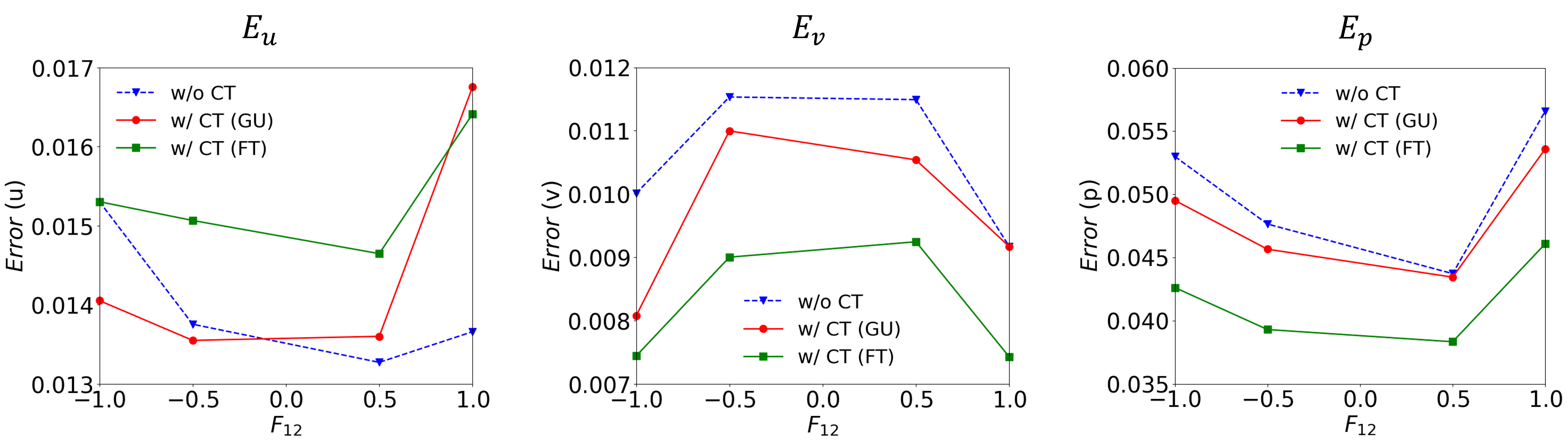}
    \caption{Comparisons of the $L_2$-norm errors of the velocity and pressure ($E_u, E_v, E_p$) for different sheared cavity shapes ($F_{12}$) between the ground truth and PINNs without/with coordinate transformation (CT), where `w/o CT', `w/ CT (GU)' and `w/ CT (FT)' denote the PINN$_{w/o-CT}$ (without CT), PINN$_{GU}$ (with CT, Glorot uniform initializer) and PINN$_{FT}$ (with CT, fine-tuned with $F_{12}^{(pre)}=0$), respectively.}
    \label{fig_def_acc}
  \end{center}
\end{figure}

Fig.~\ref{fig_def_acc} shows comparisons of the $L_2$-norm error of the PINNs against the ground truth for different cavity shapes ($F_{12}$) for PINN$_{w/o-CT}$, PINN$_{GU}$, and PINN$_{FT}$ with $F_{12}^{(pre)}=0$. The errors tend to increase as the cavity shape becomes more deformed (i.e., $F_{12}$ increases). However, the error is sufficiently small compared to the absolute value shown in Fig.~\ref{fig_s3_def_vF}. Therefore, as with the rectangular case, we conclude that coordinate transformation has little effect on the estimation accuracy.

\begin{figure}[H]
  \begin{center}
    \includegraphics[width=\linewidth]{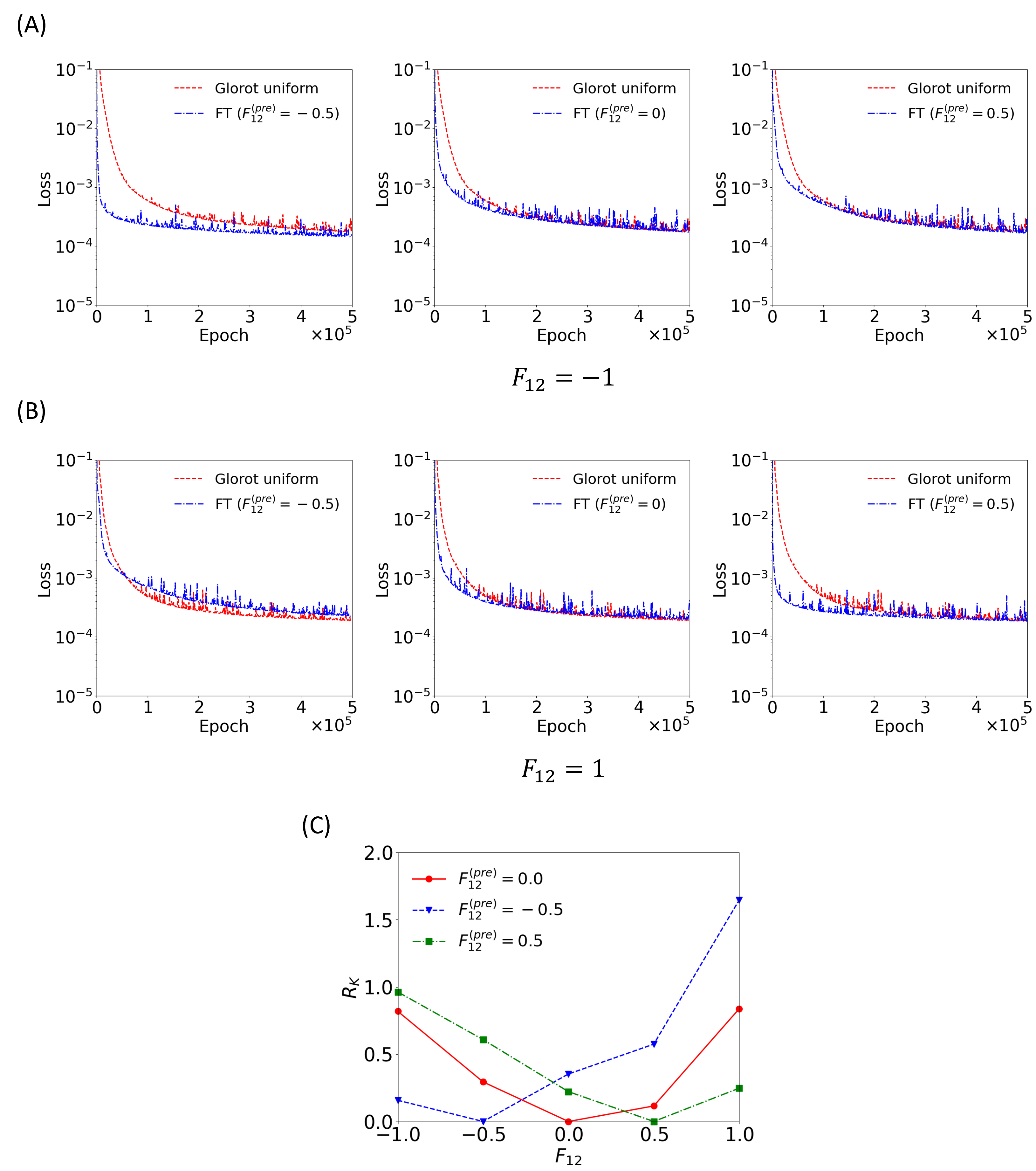}
    \caption{Comparison of the convergence behavior of fine-tuning (FT) obtained with Glorot uniform and different pre-trained conditions ($F_{12}^{(pre)}=-0.5, \ 0, \ 1$) for different shear geometries: $F_{12}=-1$ (A) and $F_{12}=1$ (B), and their convergence efficiency $R_K$ (C).}
    \label{fig_s3_def_F12}
  \end{center}
\end{figure}

We individually investigate three pre-trained DNNs to verify the capability of the fine-tuning model for different $F_{12}$. Fig.~\ref{fig_s3_def_F12} shows comparisons and convergence behaviors in the Glorot uniform and fine-tuning with $F_{12}^{(pre)}=-0.5, \ 0, \ 1$ for the sheared shapes at $F_{12}=-1$ and $1$. Compared with the randomly-initialized model using the Glorot uniform initializer, the fine-tuning model reduces the number of epochs to reach the same level of total loss except in the case of $F_{12}=1$, in which the convergence behavior degrades for the fine-tuned model with $F_{12}^{(pre)}=-0.5$. Most of the convergence efficiencies are improved with the fine-tuning model, where $R_K$ falls below $1$. Analogous to the rectangular case, a pre-trained DNN with similar geometry is more effective for improving the convergence efficiency.

\begin{figure}[H]
  \begin{center}
    \includegraphics[width=\linewidth]{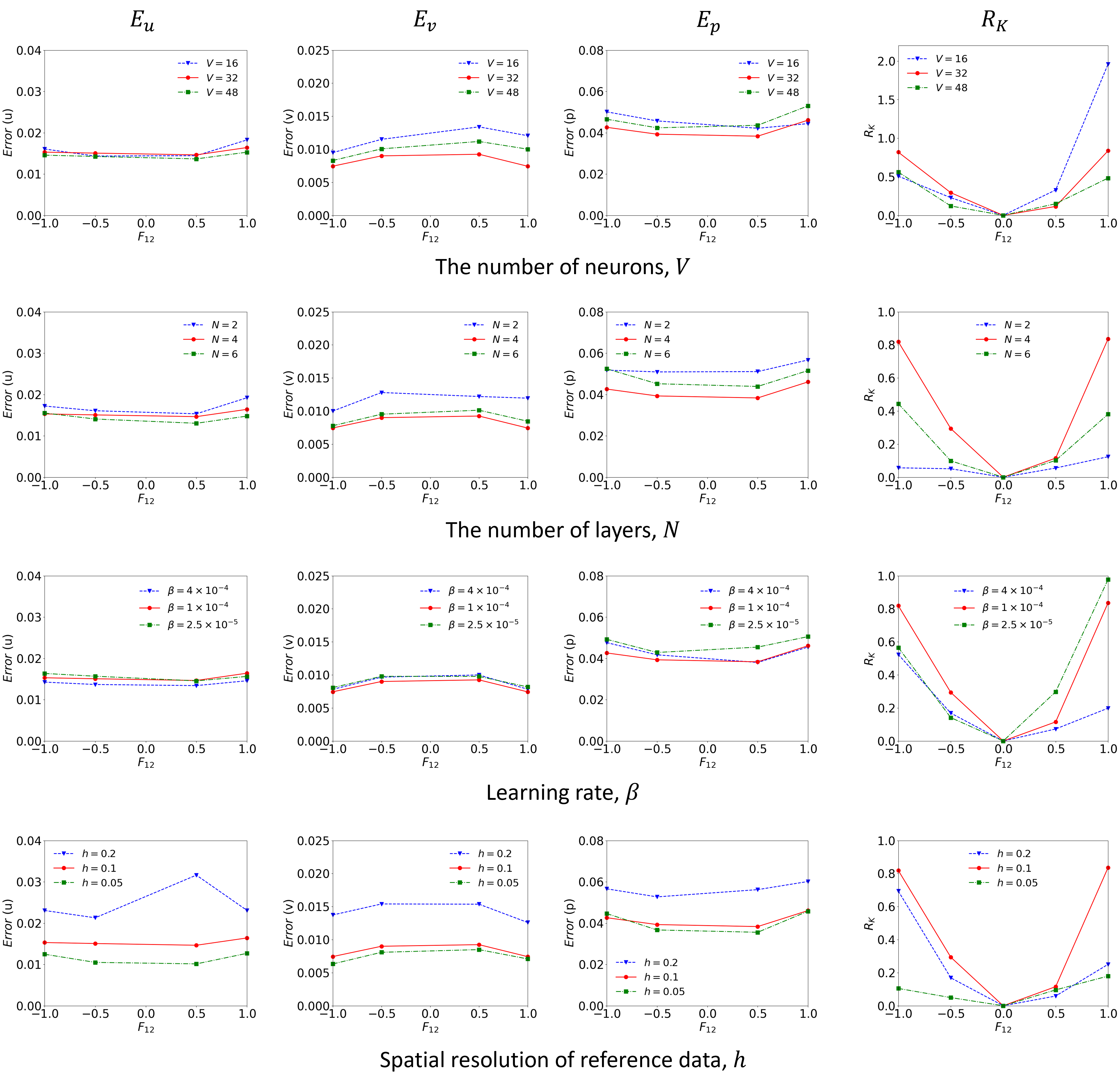}
    \caption{The influence of the hyper-parameters (number of neurons $V$, number of layers $N$, learning rate $\beta$, and spatial resolution of the reference data $h$) in the proposed fine-tuning PINNs on the $L_2$-norm error ($E_u, E_v, E_p$) relative to the ground truth and convergence efficiency $R_K$ for the cavity flow problem with shear geometries. The standard parameter set used in this study is $V=32$, $N=4$, $\beta=10^{-4}$, and $h=0.01$.}
    \label{fig_shear_param}
  \end{center}
\end{figure}

To investigate the influence of hyper-parameters in PINNs on estimation accuracy and convergence efficiency, we apply different parameter sets for the number of neurons $V$, the number of layers $N$, the learning rate $\beta$, and the spatial resolution of the reference data $h$. Again, the standard parameter sets used in this study are $V=32$, $N=4$, $\beta=10^{-4}$, and $h=0.01$. Fig.~\ref{fig_shear_param} shows comparisons of the $L_2$-norm errors ($E_u, E_v, E_p$) from the ground truth and convergence efficiency $R_K$ for different shear-deformed cavity shapes, where we apply a fine-tuned model with a pre-training condition at $F_{12}^{(pre)}=0$. Note that the value fluctuate slightly because the errors are estimated at the final epoch. Regarding the number of neurons and layers, estimation accuracy tends to decrease when the numbers are small, such as $V=16$ or $N=2$, but it shows an acceptable error when $V=32$ or $N=4$ or higher. When learning has converged sufficiently, the learning rate is not very sensitive to estimation accuracy. Regarding the resolution of the reference data, the error is large when $h=0.2$ is applied, whereas it is within a reasonable value for $h \leq 0.1$. It appears that using higher resolution reference data is desirable to capture the detailed flow field (around the cavity corner in this example). Regarding the convergence behavior $R_K$, it is difficult to clarify the trends of each parameter, but $R_K$ remains low at $h=0.05$ regardless of the deformation shape. This suggests that the proposed fine-tuning approach can be expected to dramatically improve convergence while maintaining high accuracy.

\subsection{Cavity flows with nonlinearly inflated shapes}
The last example investigates cavity flows with nonlinearly inflated shapes that undergoes the following deformation:
\begin{equation}
  \begin{cases}
    x(x^R, y^R) = x^R + 0.3 \alpha (x^R-0.5) \sin(\pi(1-y^R)), \\
    y(x^R, y^R) = y^R + 0.6 \alpha (x^R-0.5)^2 (1-y^R),
  \end{cases}
  \label{eq:nonlinear_disp}
\end{equation}
where $x^R, y^R$ is the reference coordinate for a unit square cavity, $x, y$ is the current coordinate for target inflated cavities, and $\alpha$ is the arbitrary parameter for controlling the extended cavity shape. In this example, we consider nonlinear deformation with a given displacement from a unit square cavity. We examine the proposed method by training under different conditions for $\alpha$ and $Re$.

\begin{figure}[H]
  \begin{center}
    \includegraphics[width=\linewidth]{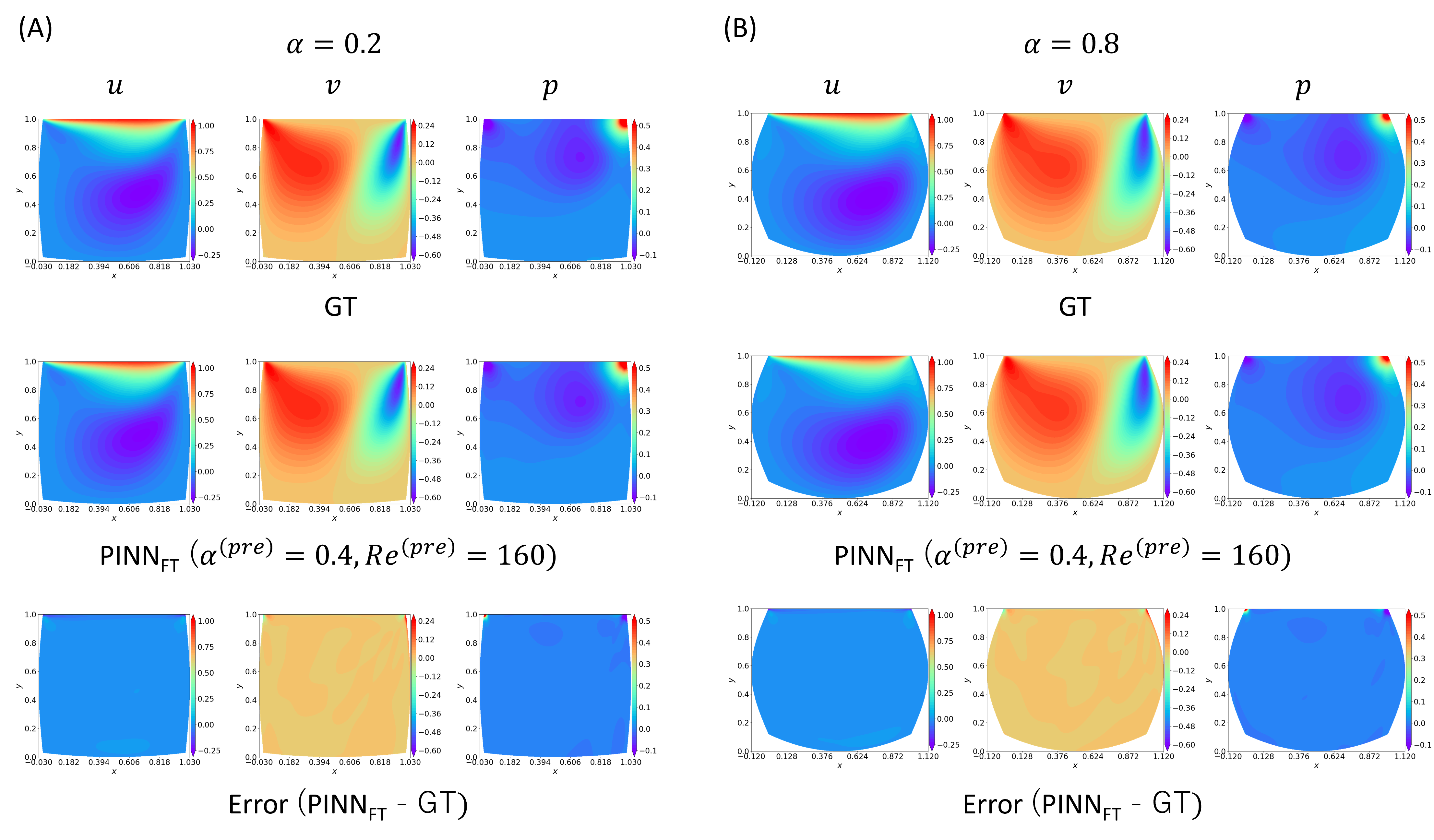}
    \caption{Comparison of velocities $u$, $v$, and pressure $p$ from the ground truth (GT) and the fine-tuned model (PINN$_{FT}$) with a pre-training condition of $\alpha^{(pre)}=0.4$ for the cavity flow problems with nonlinear inflation shape under the conditions of $\alpha=0.2, 0.8$ and $Re=160$. The errors for each are also shown.}
    \label{fig_nonlinear_map}
  \end{center}
\end{figure}

\begin{figure}[H]
  \begin{center}
    \includegraphics[width=\linewidth]{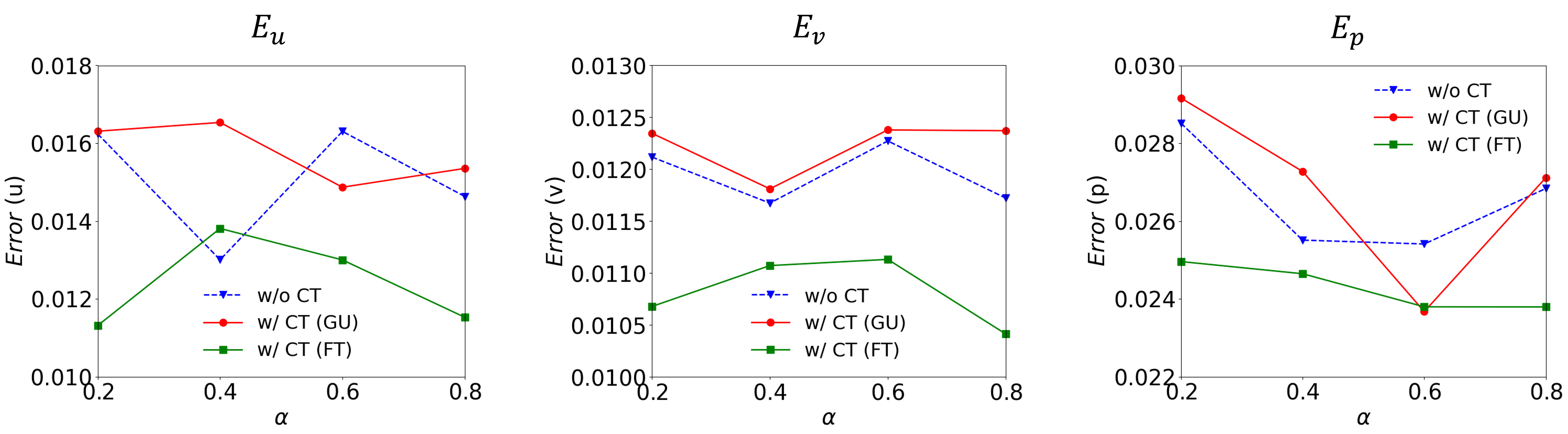}
    \caption{Comparison of the ground truth and PINNs with and without coordinate transformation (CT) for velocity and pressure $L_2$-norm errors ($E_u, E_v, E_p$) for different nonlinearly extended cavity shapes with varying $\alpha$. Here, `w/o CT', `w/ CT (GU)', and `w/ CT (FT)' denote PINN$_{w/o-CT}$ (without CT), PINN$_{GU}$ (with CT, Glorot uniform initializer), and PINN$_{FT}$ (with CT, fine-tuned with $\alpha^{(pre)}=0.4$ and $Re^{(pre)}=160$), respectively. Note that the flow field is applied with $Re=160$.}
    \label{fig_nonlinear_acc}
  \end{center}
\end{figure}

Fig.~\ref{fig_nonlinear_map} shows flow fields for the ground truth and fine-tuning results using the pre-trained condition of $\alpha^{(pre)}=0.4$ and $Re^{(pre)}=160$ for the estimation at $\alpha=0.2, 0.4$ and $Re=160$. The velocity and pressure fields of each ground truth are suitably reproduced by the proposed fine-tuning model. Again, the PINN$_{w/o-CT}$, which is based on the target deformed shapes in a current configuration (Section 2.2), is applied for investigating the effects of coordinate transformation. From the displacement fields \eqref{eq:nonlinear_disp}, the mapping of the hard constraint is defined as ${\bf u}=\mathcal{B}(\tilde{\bf u})=f_L(x,y)f_R(x,y)f_B(x,y)\tilde{\bf u}$, where
\begin{align}
  &f_L(x,y) = x + 0.15 \alpha \sin\left(\frac{\pi (1 - y)}{1 - 0.15\alpha}\right), \\
  &f_R(x,y) = 1 - x + 0.15 \alpha \sin\left(\frac{\pi (1 - y)}{1 - 0.15\alpha}\right), \\
  &f_B(x,y) = - 0.6\alpha (x-0.5)^2 + y.
\end{align}
Fig.~\ref{fig_nonlinear_acc} shows the $L_2$-norm error against the ground truth for PINN$_{w/o-CT}$, PINN$_{GU}$, and PINN$_{FT}$ with pre-training of $\alpha^{(pre)}=0.4$ and $Re^{(pre)}=160$. Comparing the absolute values shown in Fig.~\ref{fig_nonlinear_map} reveals that the errors are sufficiently small, and that the error level is slightly lower for the fine-tuned model. Furthermore, as with other shapes, coordinate transformation does not affect estimation accuracy.


In Fig.~\ref{fig_nonlinear_map}, we compare convergence efficiencies $R_K$ for different $\alpha$ and $Re$ using the fine-tuning model with different pre-trained sets $(\alpha^{(pre)}, Re^{(pre)}) = (0.4, 160)$ and $(0.6, 160)$. The reasonable convergence efficiency can be observed (i.e., $R_K<1$) and the overall trend shows that the effective improvement is achieved around the pre-trained conditions. The non-monotonic behaviors of $R_K$ are observed that would be attributed to a similarity and difference of flow fields determined by the geometry and flow feature.

\begin{figure}[H]
  \begin{center}
    \includegraphics[width=\linewidth]{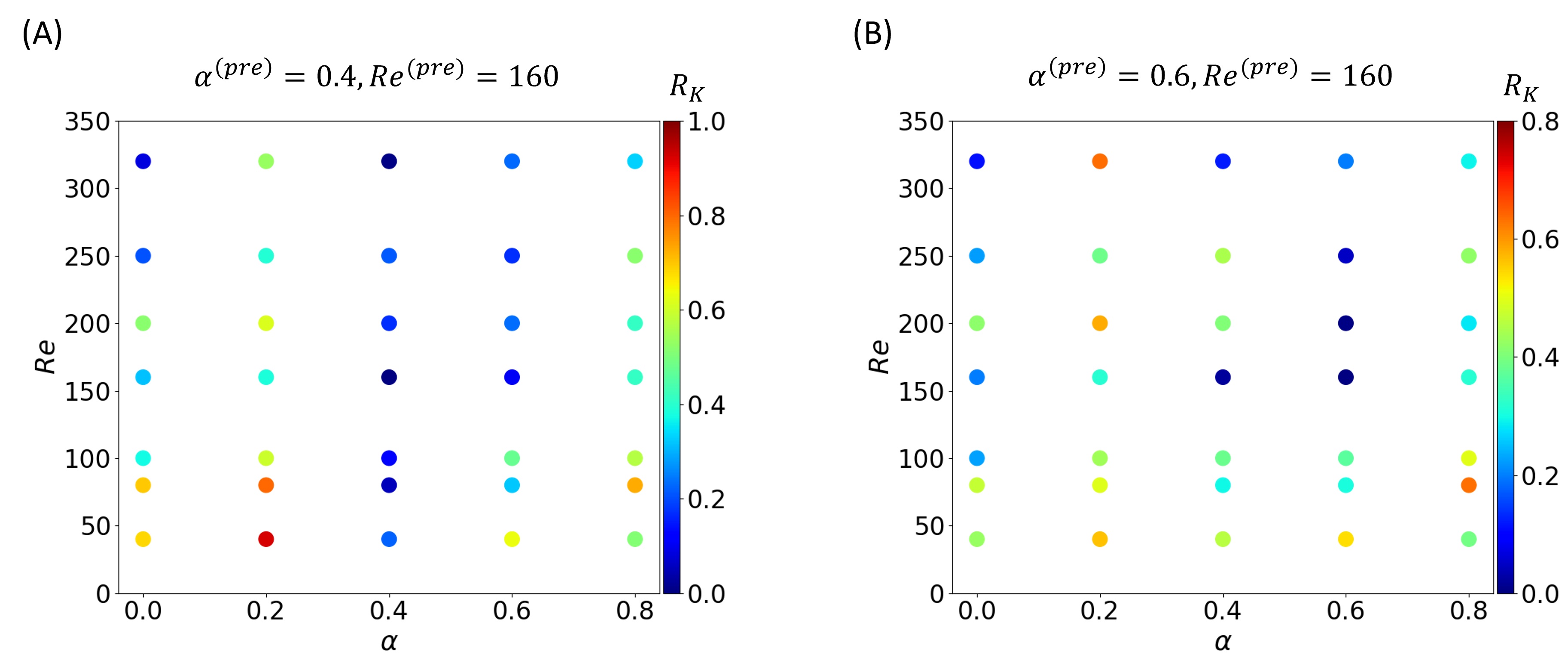}
    \caption{Convergence efficiencies $R_K$ for nonlinearly inflated cavity shapes with different $\alpha$ and $Re$ using the fine-tuning model with different pre-trained sets $(\alpha^{(pre)}, Re^{(pre)}) = (0.4, 160)$ (A) and $(0.6, 160)$ (B).}
    \label{fig_nonlinear_Rk}
  \end{center}
\end{figure}

%%% Subsection %%%
\subsection{Future outlook}
There are several aspects of future work to prepare the proposed approach for general application.

In this work, the analysis domain is simple and known in advance, i.e., the analytical form of the deformation gradient ${\bf F}$ is given. One of the goals of this research is the quantification of intra-aneurysmal blood flows, and for this purpose, the cavity-like shape is sufficient. However, patient-specific simulations necessitate more complex geometry and require a method to evaluate the deformation gradient in three-dimensional fields. One idea is to solve an elliptic problem or elastostatic problem for coordinate transformation with boundary conditions of surface displacements between the reference and target configurations \cite{Gao2021}. An adequate reference configuration is also a crucial issue because a square/cubic domain has singular corners and edges. A hemispherical shape is preferable to address the intra-aneurysmal flows. Furthermore, there is a gap between current 2D benchmarks and complex 3D scenarios. Although the proposed approach is not limited to 2D problems, further investigations using idealized and real-world 3D geometries are required.

Although this study prescribes an ideal reference configuration (i.e., square domain) for setting hard constraints of boundary conditions, more flexible treatments are required when various reference configuration will be used to increase convergence efficiency.

The efficiency of the proposed fine-tuning approach should be clarified for cases where both $Re$ and the shapes are dramatically changed. This complicates the choice of an appropriate pre-trained model considering the conditions. A potential approach is to use indices to semiautomatically determine the optimal pre-trained model from a database or a deep-learning operator combined with PINNs to consider multiple geometries and flow conditions \cite{Wang2021}.

In this study, we introduced a relative metric of convergence efficiency, $R_K$, because in a few cases, the loss at the final epoch of a fine-tuned model using pre-trained conditions with significantly different shapes was larger than that of a randomly initialized model. In practical cases, learning is sufficient if the flow field converges to a certain level, so we believe that the relationship between the absolute loss and the convergence of the flow field needs further investigation.

Theoretical evidence that the proposed fine-tuning model using a pre-trained model as a model initializer performs well in various cases, including 3D problems, has not yet been presented. Detailed analysis under simplified conditions is required, as shown in \cite{Difonzo2025} and elsewhere. It should also be noted that the convergence efficiency can be further improved by combining it with state-of-the-art convergence techniques such as adaptive scaling of loss weights \cite{Berardi2025}.

%%% Section %%%
\section{Concluding remarks}
In this study, we develop a fine-tuning approach of PINNs for lid-driven cavity flows using a pre-trained model with a given reference shape and $Re$. To address different cavity shapes, we embed the coordinate transformation for the stationary incompressible NS equations in the DNN model as a PDE loss. By performing numerical experiments using synthetic data obtained from computational fluid dynamics, we derive the following insights.

\begin{itemize}
\item A pre-trained model with the same shape but different Reynolds number can improve the convergence efficiency for square cavity flows.
\item When transforming an arbitrary target shape into a reference shape, the PINNs can reproduce the original (ground truth) flow profiles.
\item The pre-trained model effectively improves the convergence efficiency for rectangular, sheared and nonlinearly inflated shapes, and the improvement is greater when applying a cavity shape similar to the target shape. 
\end{itemize}

Previous studies have attempted various approaches to evaluate the internal flow of cerebral aneurysms using real-world data \cite{Ii2018, Funke2019, Gaidzik2019, Raissi2020, Ichimura2025}. However, challenges include low measurement accuracy due to biological noise, low spatiotemporal resolution due to measuring device limitations, and enormous computational costs. Thus, a highly accurate, high-resolution, and efficient evaluation method has yet to be established. In this regard, the proposed fine-tuning approach of PINNs is expected to shorten the learning and estimation time for patient-specific cerebral aneurysms.

%%% Section (format) %%%
\section*{CRediT authorship contribution statement}
\textbf{Ryuta Takao:} Writing -- review \& editing, Writing -- original draft, Visualization, Validation, Software, Methodology, Investigation, Formal analysis. \textbf{Satoshi Ii:} Writing -- review \& editing, Visualization, Supervision, Software, Resources, Methodology, Investigation, Funding acquisition, Data curation, Conceptualization.

%%% Section (format) %%%
\section*{Declaration of competing interest}
The authors declare that they have no known competing financial interests or personal relationships that could have appeared to influence the work reported in this paper. 

%%% Section (format) %%%
\section*{Data availability}
Data will be made available on request.

%%% Section (format) %%%
\section*{Acknowledgements}
The authors acknowledge Tsubasa Ichimura (Tokyo Metropolitan University) and Kakeru Ueda (Institute of Science Tokyo) for fruitful discussion of the inverse problem of cavity flows. This work was supported by the MEXT Program for Promoting Researches on the Supercomputer Fugaku (Development of human digital twins for cerebral circulation using Fugaku, JPMXP1020230118) and used computational resources of the supercomputer Fugaku provided by the RIKEN Center for Computational Science (project ID: hp230208, hp240220, hp250236). The authors thank Irina Entin, M. Eng., from Edanz (https://jp.edanz.com/ac) for editing a draft of this manuscript.

%%% Section (format) %%%
\appendix

%%% Subsection %%%
\section{Incompressible NS equations with coordinate transformation}
Using the deformation gradient tensor ${\bf F}=\partial{\bf x}/\partial{\bf x}^R=(\nabla^R{\bf x})^\top$ defined as
\begin{align}
  {\bf F} = F_{ij}{\bf e}_i{\bf e}^R_j
   = \frac{\partial x_i}{\partial x^R_j}{\bf e}_i{\bf e}^R_j,
\end{align}
where ${\bf e}_i$ and ${\bf e}^R_j$ are the unit basis vectors in the current and reference configurations and the Einstein summation convention is used, the velocity gradient tensor ${\bf L}=\nabla{\bf u}$ can be written as
\begin{align}
  {\bf L} = L_{ij}{\bf e}_i{\bf e}_j 
  = \frac{\partial u_j}{\partial x_i}{\bf e}_i{\bf e}_j
  = G_{li}\frac{\partial F_{jk}u^R_k}{\partial x^R_l}{\bf e}_i{\bf e}_j
  = G_{li}\left(\frac{\partial u^R_k}{\partial x^R_l}F_{jk}
  + \frac{\partial F_{jk}}{\partial x^R_l}u^R_k
  \right){\bf e}_i{\bf e}_j,
\end{align}
where ${\bf G}=G_{ij}{\bf e}^R_i{\bf e}_j=(\partial x^R_i/\partial x_j){\bf e}^R_i{\bf e}_j={\bf F}^{-1}$.

Thus, Eqs. \eqref{eq:continuity}, \eqref{eq:momentum} and \eqref{eq:viscous} are rewritten as
\begin{align}
  & \nabla\cdot{\bf u} = L_{ii} = 0, \\
  & {\bf u}\cdot\nabla{\bf u} + \nabla{p} - \nabla\cdot{\bm \tau}
  = \left(
    F_{ik} u^R_k L_{ij}
    + G_{ij}\frac{\partial p}{\partial x^R_i}
    - G_{ik}\frac{\partial \tau_{kj}}{\partial x^R_i}
  \right){\bf e}_j
  = 0, \\
  & {\bm \tau} - \frac{1}{Re} (\nabla{\bf u} + \nabla{\bf u}^\top)
  = \left(
    \tau_{ij} - \frac{1}{Re} (L_{ij}+L_{ji})
  \right){\bf e}_i{\bf e}_j
  = 0.
\end{align}
%

%%% Subsection %%%
\section{Loss definition in the PINNs with coordinate transformation}
The data loss of the velocity mismatch $\mathcal{L}_{data}$ is defined as
\begin{align}
  \mathcal{L}_{data} = \sum_{i=1}^{N_{data}} ({\bf u}^R|_i - {\bf U}^R|_i)^2,
\end{align}
where $N_{data}$ is the number of reference data, $|_i$ denotes the value at ${\bf x}^R_i$, e.g., ${\bf u}^R|_i={\bf u}^R({\bf x}^R_i)$, and ${\bf U}^R|_i$ is the reference (observation) velocity mapped to the reference coordinate ${\bf x}^R|_i$.

The PDE loss $\mathcal{L}_{PDE}$ is defined as
\begin{align}
  \begin{split}
  \mathcal{L}_{PDE} = 
  & \sum_{i=1}^{N_{PDE}} {\rm tr}({\bf L}|_i)^2 \\
  & + \sum_{i=1}^{N_{PDE}} ({\bf F}|_i\cdot{\bf u}^R|_i\cdot{\bf L}|_i - {\bf G}^\top|_i\cdot (-\nabla^R{p}|_i + \nabla^R\cdot{\bm \tau}|_i))^2 \\
  & + \sum_{i=1}^{N_{PDE}} 
  \left(
    {\bm \tau}|_i - \frac{1}{Re}({\bf L}|_i + {\bf L}^\top|_i)
    \right)^2.
  \end{split}
\end{align}
where $N_{PDE}$ is the number of PDE points and
\begin{align}
  {\bf L}|_i = {\bf G}^\top|_i \cdot (\nabla^R{\bf u}^R|_i\cdot{\bf F}^\top|_i + \nabla^R{\bf F}|_i\cdot{\bf u}^R|_i).
\end{align}
%

%%% Reference %%%
\bibliographystyle{elsarticle-num}
\bibliography{ref}

\end{document}